\documentclass[a4paper,fleqn,usenatbib]{mnras}

\usepackage{newtxtext,newtxmath}

\usepackage[T1]{fontenc}
\usepackage{ae,aecompl}

\usepackage{deluxetable}
\usepackage{graphicx}	
\usepackage{amsmath}	
\usepackage{amssymb}	

\title[Origin and properties of prolate galaxies]{The origin and properties of massive prolate galaxies in the Illustris simulation}

\author[Hongyu~Li et al.]{Hongyu~Li$^{1,2}$\thanks{E-mail: hyli@nao.cas.cn}, Shude~Mao$^{3,1,4}$,  Eric Emsellem$^{5,6}$,
 Dandan Xu$^{7}$, Volker Springel$^{7,8}$ \and and Davor Krajnovi{\'c}$^{9}$
\vspace*{0.2cm}\ \\
$^{1}$National Astronomical Observatories, Chinese Academy of Sciences, 20A Datun Road, Chaoyang District, Beijing 100012, China\\
$^{2}$University of Chinese Academy of Sciences, Beijing 100049, China\\
$^{3}$Physics Department and Tsinghua Centre for Astrophysics, Tsinghua University, Beijing 100084, China\\ 
$^{4}$Jodrell Bank Centre for Astrophysics, School of Physics and Astronomy, The University of Manchester, Oxford Road,\\ Manchester M13 9PL, UK\\
$^{5}$Universit\'e Lyon 1, Observatoire de Lyon, Centre de Recherche
Astrophysique de Lyon and Ecole Normale Sup\'erieure de Lyon,\\ 9 avenue 
Charles Andr\'e, F-69230 Saint-Genis Laval, France\\
$^{6}$European Southern Observatory, Karl-Schwarzschild-Str. 2, 85748 Garching, Germany\\
$^{7}$Heidelberg Institute for Theoretical Studies, Schloss-Wolfsbrunnenweg 35, 69118 Heidelberg, Germany\\
$^{8}$Zentrum f\"{u}r Astronomie der Universit\"{a}t Heidelberg, ARI, M\"onchhofstr. 12-14, 69120 Heidelberg, Germany \\
$^{9}$Leibniz-Institut f\"ur Astrophysik Potsdam (AIP), An der Sternwarte 16, D-14482 Potsdam, Germany\\
}

\date{Accepted 2017 September 10. Received 2017 August 08; in original form 2017 June 12}

\pubyear{2017}

\begin{document}
\label{firstpage}
\pagerange{\pageref{firstpage}--\pageref{lastpage}}
\maketitle

\begin{abstract}
  We study galaxy shapes in the Illustris cosmological hydrodynamic simulation. We find that massive galaxies have a higher probability of being prolate. For galaxies with stellar mass larger than $10^{11}\rm M_{\odot}$, 35 out of total 839 galaxies are prolate. For 21 galaxies with stellar mass larger than $10^{12}\rm M_{\odot}$, 9 are prolate, 4 are triaxial while the others are close to being oblate. There are almost no prolate galaxies with stellar mass smaller than $3\times10^{11}\rm M_{\odot}$. We check the merger history of the prolate galaxies, and find that they are formed by major dry mergers. All the prolate galaxies have at least one such merger, with most having mass ratios between $1:1$ and $1:3$. The gas fraction (gas mass to total baryon mass) of the progenitors is 0-3\% percent for nearly all these mergers, except for one whose second progenitor contains $\sim 15\%$ gas mass, while its main progenitor still contains less than $5\%$. For the 35 massive prolate galaxies that we find, 18 of them have minor axis rotation, and their angular momenta mostly come from the spin angular momenta of the progenitors (usually that of the main progenitor). We analyse the merger orbits of these prolate galaxies and find that most of them experienced a nearly radial merger orbit. Oblate galaxies with major dry mergers can have either radial or circular merger orbits. We further discuss various properties of these prolate galaxies, such as spin parameter $\lambda_{\rm R}$, spherical anisotropy parameter $\beta$, dark matter fraction, as well as inner density slopes for the stellar, dark matter and total mass distributions.

\end{abstract}




\section{Introduction}
Ellipses provide a good approximation of 2-dimensional isophotes of early-type galaxies. The 3-dimensional shapes of these galaxies, however, are degenerate with viewing angle. One can construct the same 2-dimensional surface brightness with an oblate, prolate or triaxial intrinsic shape \citep{Rybicki1987,Franx1988,Monnet1992,Ryden1992,Emsellem1994,Statler1994,
Tremblay1996,Bosch1997}. Combined with spatial resolved kinematic data, e.g. $\rm ATLAS^{3D}$ \citep{Cappellari2011}, CALIFA \citep{Sanchez2012}, MASSIVE \citep{Ma2014}, SAMI \citep{Bryant2015} and MaNGA \citep{Bundy2015}, one can reduce the degeneracy by measuring the angle between kinematic and photometry axis and constrain the intrinsic shape of early-type galaxies.  Such studies show that regularly rotating early-type galaxies are mostly oblate or slightly triaxial \citep{Krajnovic2011, Weijmans2014, Fogarty2015, Cappellari2016}. But massive ellipticals, which rotate slowly \citep{Illingworth1977, Emsellem2011}, tend to have large misalignments between kinematic and photometric axes and show even minor axis rotation (rotation is about the photometric major axis), like in NGC 3923 \citep{Carter1998}, M87 \citep{Emsellem2014} and the galaxies from the MUSE Most Massive Galaxies (M3G) survey (Krajnovi{\'c} et al. 2017, in preparation), CALIFA survey \citep{Tsatsi2017} and MaNGA survey (Li et al. 2017, in preparation). These studies show that more massive galaxies are triaxial or prolate and likely have different formation scenarios.

\citet{Jesseit2005,Jesseit2009} studied the orbital parameters, the intrinsic shapes and the
kinematic misalignments of the binary mergers of \citet{Naab2003}, and found that some merger
remnants are prolate with minor-axis rotation. 
\citet{Rodriguez-Gomez2015} found that collisionless equal-mass merger of disk galaxies could 
produce prolate systems with minor-axis rotation. Similar results are obtained for gas-poor
mergers in the hydrodynamic simulation in \citet{Moody2014}.
In addition, pure dark matter simulations  (e.g. \citealt{Jing2002}) also produce dark halos
that are on average prolate, which is thought to be the consequence of repeated mergers of
dispersion supported systems. The dark halos in the hydrodynamic binary merger simulation in
\citet{Novak2006} also produce prolate or triaxial shape.
The ever increasing computational power allows the constructions of more and more realistic physical models based on cosmological simulations. The current state-of-the-art cosmological simulations, e.g. Illustris \citep{Vogelsberger2014a, Vogelsberger2014b, Genel2014} and EAGLE \citep{Schaye2015},  can well reproduce basic observational features of our Universe, and provide us with realistic galaxy samples and their evolution histories. In \citet[figure 10]{Li2016}, we found that more massive galaxies tend to be prolate in the Illustris simulation. This is also true for the most massive galaxies in the EAGLE simulation (\citealt[figure 7]{Schaller2015}; \citealt{Velliscig2015}).
\citet{Naab2014} also found in cosmological zoom simulations that some slow rotators formed 
by major dry mergers (Class E) could have minor-axis rotation.
It could help us to better understand the galaxy evolution processes if we can find the formation mechanisms of these massive prolate galaxies.  

In this paper, we study the galaxy shapes and their evolution in the Illustris simulation. 
We focus on prolate galaxies and their evolution histories and properties. The structure of this paper is as follows. In Section~\ref{sec:simulation_and_methods}, we introduce the simulation data and the methods that we use. In Section~\ref{sec:results}, we show our results concerning the mass dependence of galaxy shapes (Section~\ref{sec:shape_on_mass}), the merger histories  (Section~\ref{sec:merger_history}), merger mass ratios (Section~\ref{sec:mass_ratio}) and merger orbits (Section~\ref{sec:merger_orbit}), the origin of the minor axis rotations (Section~\ref{sec:minor_axis_rotation}) and general properties of the massive prolate galaxies (Section~\ref{sec:properties}). In Section~\ref{sec:conclusions}, we summarize and give our conclusions.
  
\section{Simulations and methods}
\label{sec:simulation_and_methods}

\subsection{The Illustris simulation}
\label{sec:illustris} 
The Illustris project \citep{Vogelsberger2014a, Vogelsberger2014b, Genel2014} comprises a suite of cosmological hydrodynamic simulations carried out with the moving mesh code {\small AREPO} \citep{springel2010}. The hydrodynamical simulation follows the evolution of the baryon component using a number of sophisticated (in part subgrid) models for the galaxy formation physics \citep{Vogelsberger2013}. The Illustris simulation reproduces various observational results, such as cosmic star formation rate density, mass-size relation \citep[figure 5]{Xu2017}, galaxy luminosity function and Tully-Fisher relation etc. The galaxy morphology type fractions as a function of stellar mass and environment also roughly agree with observations \citep{Vogelsberger2014b, Snyder2015}.

In this work, we use the largest simulation (Illustris-1) of the Illustris project which contains $1820^3$ dark matter particles and approximately $1820^3$ gas cells or stellar particles. The simulation follows the evolution of the universe in a periodic box of $106.5\: \rm Mpc$ on a side, from $z=127$ to $z=0$. The softening lengths for the dark matter and baryon components are $1420$ $\rm pc$ and $710$ $\rm pc$ respectively. The cosmological parameters adopted in the simulations are $\Omega_m=0.2726$, $\Omega_L=0.7274$, $\sigma_8=0.809$, $h=0.704$ and $n_{\rm s}=0.963$ \citep{Vogelsberger2014a}. The galaxy's particle cutout files, merger trees and catalogued galaxy properties that we use are from the Illustris public data release\footnote{\url{http://www.illustris-project.org}} \citep{Nelson2015,Xu2017}.

\subsection{Sample selection and shape measurement} \label{sec:sample} 
\begin{figure}
\includegraphics[width=\columnwidth]{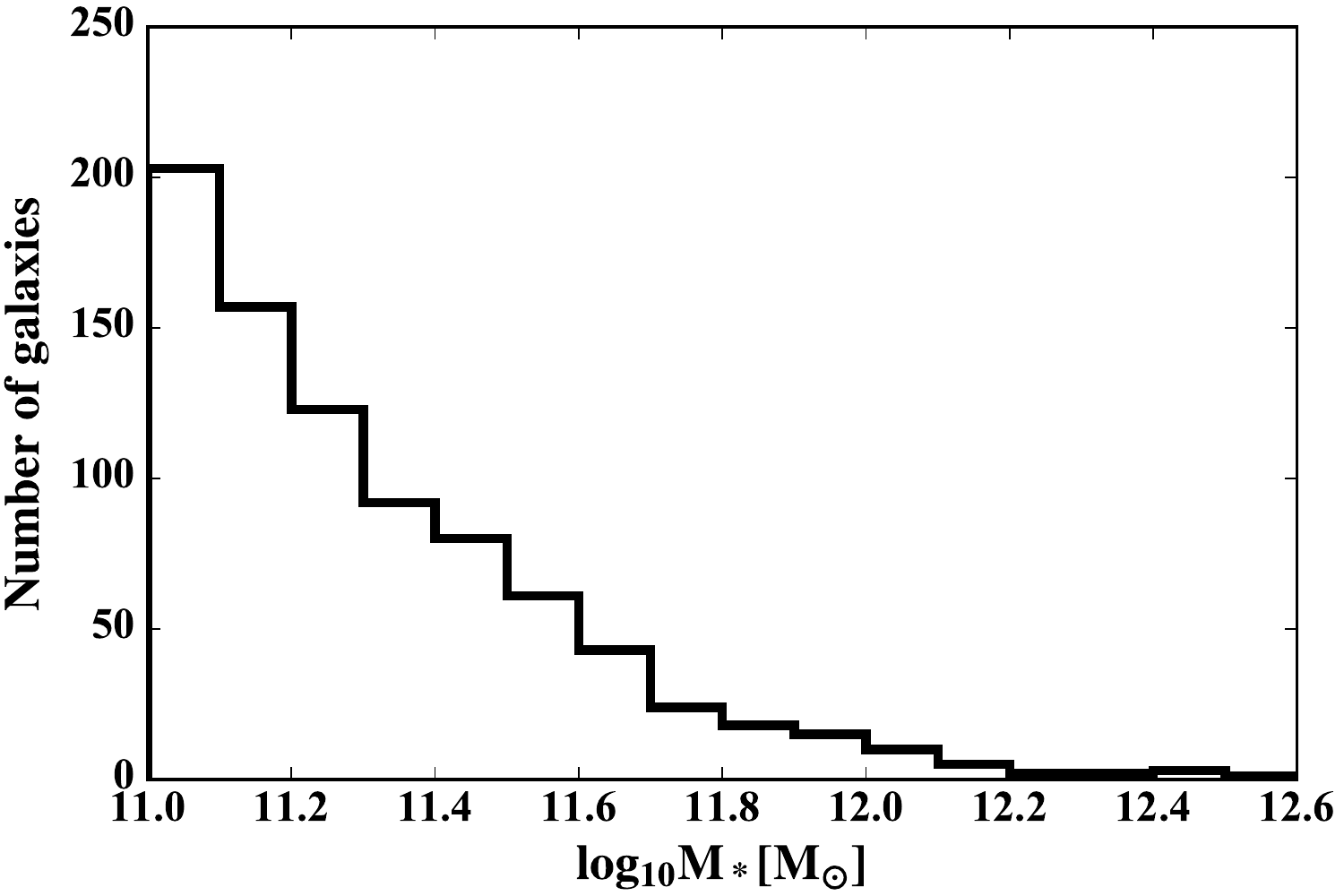}
    \caption{The stellar mass distribution of our selected sample from snapshot 135 ($z=0$). In total, there are 839 galaxies.}
    \label{fig:mass_distribution}
\end{figure}

We select our sample at redshift $z=0$ (snapshot 135) by stellar mass and light profile S{\`e}rsic index \citep{sersic1963}. The stellar mass of a galaxy is provided by the {\small SUBFIND} catalogue (for more details see \citealt{Nelson2015}, and the {\small SUBFIND} algorithm, \citealt{Springel2001}). The measurements of the S{\`e}rsic index are described in \citet{Xu2017}. Galaxies with $\rm \log M^*>11.0$ and $n_{\rm S{\`e}rsic} > 2.0$ are selected: the limit on stellar mass ensures all the simulated galaxies have enough particles to accurately measure their shapes. The limit on the S{\`e}rsic index allows us to exclude late type galaxies (97 in total). These criteria result in 839 galaxies in our sample. The stellar mass distribution is shown in Fig.~\ref{fig:mass_distribution}. Note that all galaxies are named by their {\small SUBFIND}-ID (e.g. subhalo210738), which is unique in one snapshot, but the same galaxy may have a different ID at different output times.

\begin{figure*}
\includegraphics[width=\textwidth]{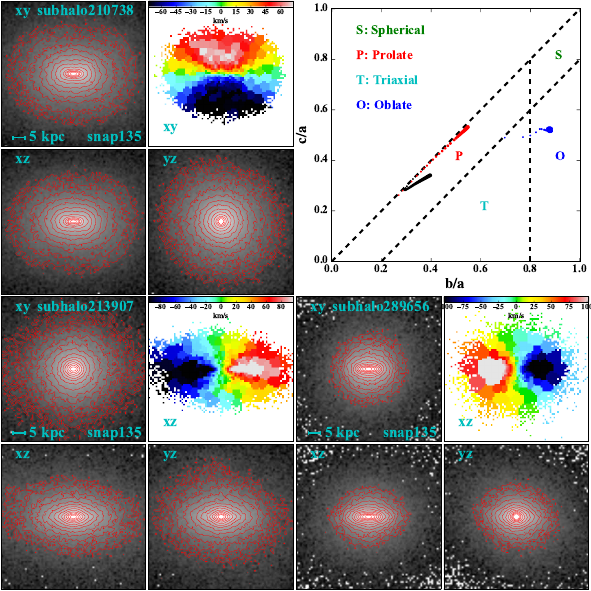}
    \caption{Upper left four sub-panels: images (grey-scale), isophotes (red-contour) in three projections and line-of-sight velocity map (color-scale) in one particular projection of a prolate galaxy. Contours are shown every 0.5 magnitude from the brightest pixel in each image to the pixels 7 magnitude lower. $x$, $y$ and $z$  correspond to axes $a$, $b$ and $c$ (see text for definition). Throughout the paper, the coordinate system for all images and velocity maps are right-hand coordinate systems. For the velocity maps, only pixels with enough S/N are plotted; red colours indicate positive velocities and blue ones negative velocities. Lower-left four sub-panels: images, isophotes in three projections and line-of-sight velocity map in one particular projection of an oblate galaxy. Upper right panel: Criteria for galaxy shape measurements according to the axis ratios $b/a$ and $c/a$. The green letter S marks the spherical galaxy region; the red letter P indicates the prolate galaxy region; the cyan letter T represents the triaxial galaxy region, and the blue letter O gives the oblate galaxy region. Red and blue dots show the axis ratios of the prolate and the oblate galaxies shown in the left panels. The axis ratios are measured from the centre to the outer part of the galaxy. The smaller the radius, the smaller the dot size. The black dots show the axis ratio distribution of a strongly barred galaxy, which can be misidentified as a prolate galaxy and whose images, isophotes in three projections and line-of-sight velocity map in one particular projection are given in the four sub-panels in the lower right.}
     \label{fig:bar_effect}
\end{figure*}
Assuming that a galaxy can be represented by ellipsoids of axis lengths $a\geq b\ge c$, the 
axis ratios (i.e. the shape) $p = b/a$ and $q =c/a$ can be measured from its stellar particles 
using the reduced inertia tensor method \citep{Allgood2006}. The tensor is defined as
\begin{equation}
I_{i,j}=\sum_{k\in \mathcal{V}} \frac{x_i^{(k)}x_j^{(k)} } {r_k^2},
\end{equation}
where  $r_k = \sqrt{x^2_k +y_k^2/p^2 +z_k^2/q^2}$ is the elliptical distance measured from 
the centre of the galaxy to the $k$-th particle, $x_i^{(k)}$ is the $i$-th coordinate of the $k$-th 
particle and $\mathcal{V}$ is the set of particles of interest. 
We calculate $p$ and $q$ iteratively. $p$ and $q$ are initially set to 
1, and $\mathcal{V}$ contains the particles with $r_k$ smaller than some radius $R$. 
In each iteration, we first calculate the tensor using the particles in $\mathcal{V}$,
and reset the value of $p$ and $q$ as the ratio of $\sqrt{\lambda_{i}}$, where $\lambda_{i}$ 
are the eigenvalues of the tensor $I$. Then we redefine the set $\mathcal{V}$  using the updated
values of $p$ and $q$.  We keep iterating until the values of $p$ and $q$ converge.
The directions of the principal axes are given by the corresponding final eigenvectors.
For every galaxy in our sample, we measure their axis ratios at different radii between
1.5 kpc and 2.5 $r_{h}^{*}$, where $r_{h}^{*}$ is the 3-dimensional spherical radius
including half of the total stellar mass.

\subsection{Selecting prolate galaxies}
\label{sec:select_prolate} 

\begin{figure}
\includegraphics[width=\columnwidth]{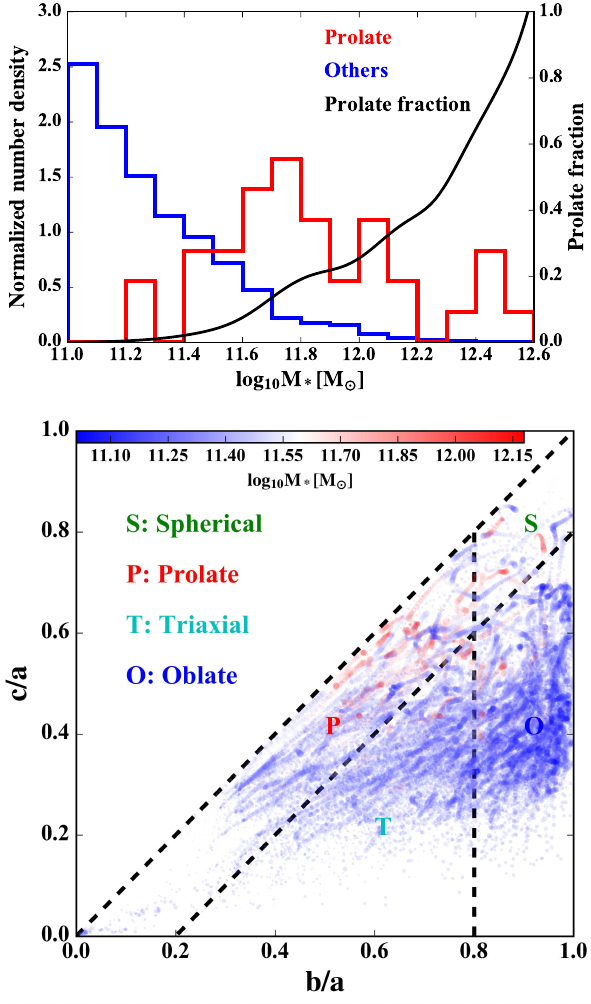}
    \caption{Top: the normalised stellar mass distribution for 35 prolate galaxies (red) and the other 804 galaxies (blue). The fraction of prolate galaxies is shown by the black solid line, which is smoothed using a Gaussian kernel. Bottom: axis ratios at different radii for all the 839 galaxies, colour coded by stellar mass. The other symbols are the same as in Fig.~\ref{fig:bar_effect}.}
    \label{fig:mass_dependence}
\end{figure}

We define galaxy shape according to these axis ratios, as shown in Fig.~\ref{fig:bar_effect}, where we show the images in three projections and line-of-sight velocity map in one particular projection of a prolate, an oblate and a strongly-barred galaxy, and their relative locations on the axis ratio diagram. Prolate galaxies are galaxies with axis ratio $c/b$ close to one (i.e. $b/a$ close to $c/a$), i.e., approximately axial symmetric about their longest axis $a$. We define the galaxies with $b/a-c/a<0.2$ and $b/a<0.8$ as prolate galaxies, as shown by the dashed lines in Fig.~\ref{fig:bar_effect}. After defining the shape according to the axis ratios, we visually check their images and velocity maps in three projections to find out the misidentified galaxies. The images and velocity maps are constructed with stellar particles. We project the particles onto a 2-dimensional grid, which has a resolution of 0.75 kpc/pixel. The flux in each image pixel (i.e. grid cell) is defined as the stellar mass within the grid cell (i.e. assuming the stellar mass-to-light ratio equals to 1). Then the grid cells are Voronoi binned \citep{Cappellari2003} to $\sim 1000$ stellar particles per bin. We then calculate for each Voronoi cell the stellar mass weighted mean velocity to obtain the velocity maps.
The visual verification process is important
because strongly barred galaxies tend to be located in the prolate region, although they have a fast rotating disk component. In the lower right of Fig.~\ref{fig:bar_effect}, we show a bar galaxy as an example, the axis ratio of
which mimics that of a prolate galaxy. In addition, other weakly barred galaxies are usually located in the triaxial region because their axis ratios are contributed by two components -- a prolate like bar in the inner part and a disk in the outer part. Mergers and close galaxy pairs will also affect the shape measurement. In these cases, the axis ratios at different radii usually have large scatters.

\section{Results}
\label{sec:results} 
\subsection{Shape dependence on galaxy stellar mass}
\label{sec:shape_on_mass}

As discussed in Section~\ref{sec:select_prolate}, we measure the axis ratios of every galaxy at radii between 1.5 kpc to 2.5 $r_{h}^{*}$. Combined with visual classification, we find 35 prolate galaxies out of a total of 839 galaxies. We compare the stellar mass distribution of these prolate galaxies against those of other galaxies, as shown in Fig.~\ref{fig:mass_dependence}.  As can be seen, the prolate galaxies are more massive than the other galaxies in the sample, and become dominating after $\log M^*>11.6$. This agrees with \citet[figure 10]{Li2016}. In \citet{Li2016}, there are prolate galaxies at the low mass end, which is due to the contamination from barred galaxies as shown in Section~\ref{sec:select_prolate}. In the bottom panel of Fig.~\ref{fig:mass_dependence}, we plot the axis ratios $b/a$ vs. $c/a$ measured at different radii (represented by dot size) for every galaxy. The stellar masses are shown by different colours. Many massive galaxies are located in the prolate region, while the oblate region is dominated by galaxies with lower stellar masses. This thus exhibits a strong dependence of galaxy shape on stellar mass. The galaxies in the lower left corner of the prolate region and triaxial region are mostly barred galaxies or merging galaxies.

\subsection{Merger history}
\label{sec:merger_history}

\begin{figure*}
\includegraphics[width=\textwidth]{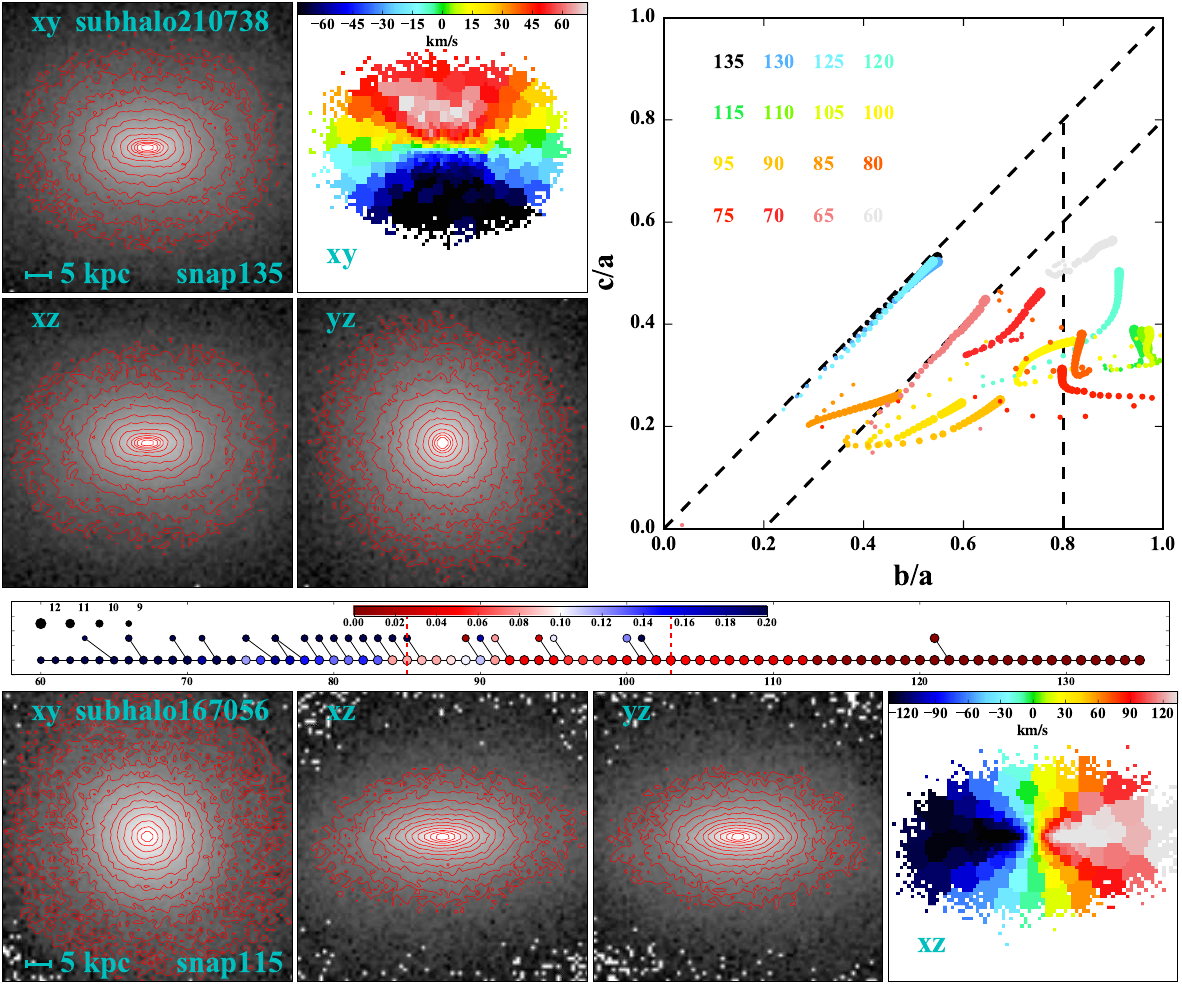}
    \caption{A prolate galaxy example, subhalo210738, with a major dry merger at snapshot 122. 
     The upper left panels show the images, isophotes projected on the $xy$, $xz$ and $yz$ planes, and line-of-sight velocity map on the $xy$ plane, where the $x$ axis is aligned with the longest axis $a$, and the $z$ axis is aligned with the shortest axis $c$. The upper right panel shows the axis ratio evolution of this galaxy. Different colours represent the shape of its progenitors at different snapshots, which is listed in the upper left of this panel. The dot size represents the radius within which the shapes are calculated. The smaller the dots, the smaller the radii. Others labels are the same as in Fig.~\ref{fig:bar_effect}. The middle panel shows the merger tree between snapshot 60 to snapshot 135 of this galaxy. The size of the dot represents the stellar mass, colour represents the gas fraction. Vertical dashed lines show the redshifts equal to 1.0 and 0.5, respectively. Mergers with mass ratio smaller than 1:100 are not shown on this plot. The bottom panels show the images, isophotes and line-of-sight velocity map of the main progenitor (oblate) of this galaxy at snapshot 115.}
    \label{fig:subhalo210738}
\end{figure*}

\begin{figure*}
\includegraphics[width=\textwidth]{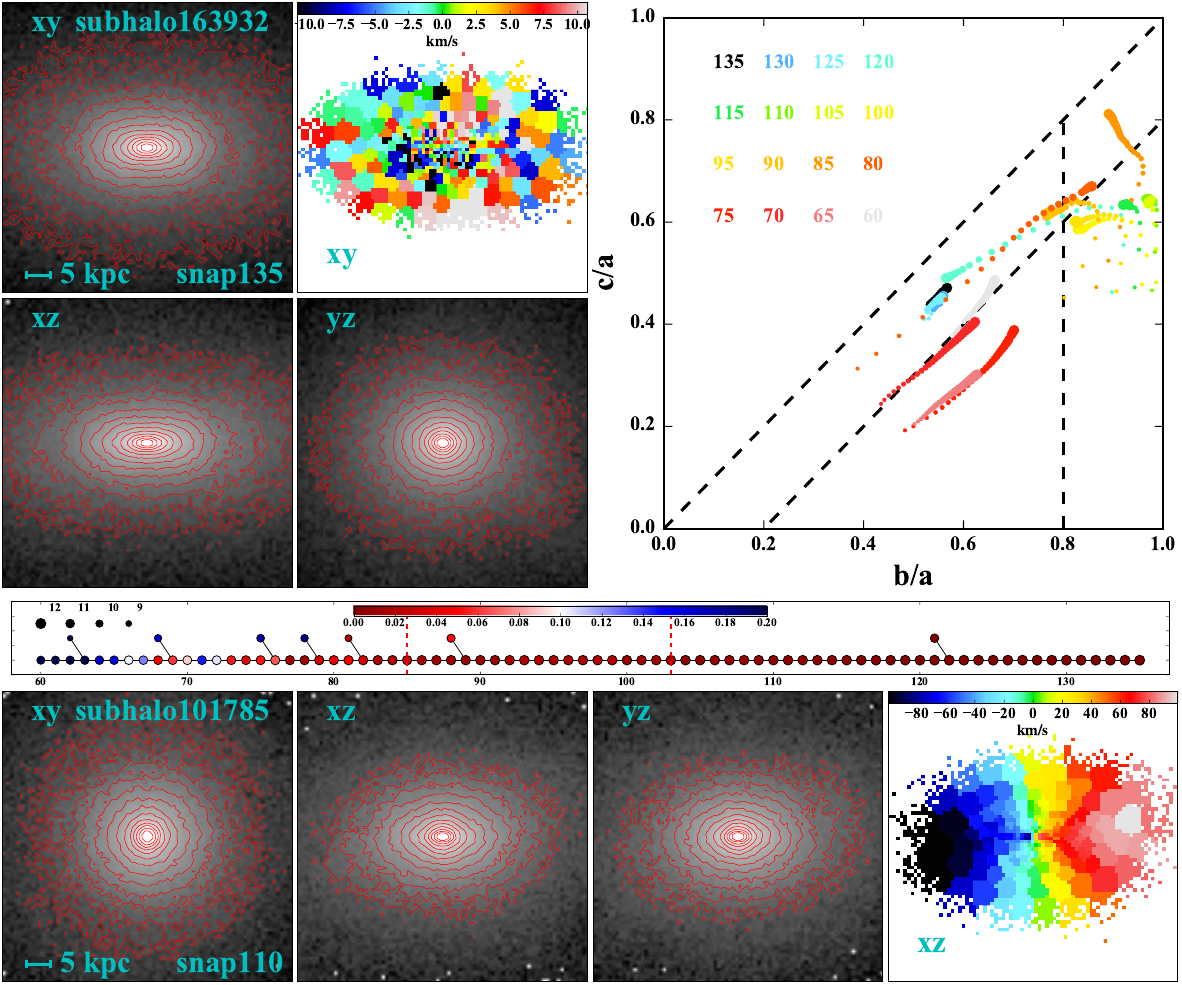}
    \caption{Another example for a prolate galaxy, subhalo163932, with a major dry merger at snapshot 122. Labels
     are similar to Fig.~\ref{fig:subhalo210738}.}
    \label{fig:subhalo163932}    
 \end{figure*}
 
 In order to understand the formation mechanism of these massive prolate galaxies, we examine their merger histories. The merger tree we use is the {\small SUBLINK} tree \citep{Rodriguez-Gomez2015} provided by the Illustris team. In a {\small SUBLINK} tree, the progenitor with the ``most massive history'' \citep{Rodriguez-Gomez2015} is defined as the main progenitor. In contrast, we define the progenitor (from the previous snapshot) with the most massive stellar mass as progenitor1 (i.e. the main progenitor) and the one with the second most massive stellar mass as progenitor2. This is because we believe the stellar mass has stronger effects on the shape of the merger remnants, while the mass history gives more information about the full evolution history up to high redshift, which does not have an equally direct impact on the merger remnant. Despite the differences in our definition, we note that the main progenitor branch of the merger trees according to our definition (walking back in time following along the main progenitor) are the same as in {\small SUBLINK} for most galaxies.

After having the merger histories of these galaxies, we use the method described in Section~\ref{sec:sample} to measure the shape of their main progenitors at different radii. The  progenitors are selected every 5 snapshots from snapshot 60 to 135. In Fig.~\ref{fig:subhalo210738} and Fig.~\ref{fig:subhalo163932}, we show the photometric and kinematic properties of two prolate examples and their main progenitors, together with the merger trees and the shape evolutions. 

For the first example, subhalo210738 (see Fig.~\ref{fig:subhalo210738}), the images, isophotes and velocity map are shown in the upper left panels. From the isophotes projected in three directions ($x$, $y$, $z$ corresponding to axis $a$, $b$ and $c$), it is clearly seen that axis $a$ and $b$ are nearly the same, and the shape is prolate. The  line-of-sight velocity map shows that this galaxy also has minor axis rotation with velocities around 80 ${\rm km\,s^{-1}}$. In the upper right panel, we show the shape evolution of this galaxy. The shapes for different snapshots are shown with different colours and labelled by the coloured numbers in the upper left panel. As can be seen, the galaxy is oblate between snapshot 105 and 115 (green colours), starts getting affected by its companion at  snapshot 120, and becomes prolate after the merger at snapshot 125. Then the shape remains unchanged until redshift $z=0$ (at snapshot 135). In the middle panel, we show the merger history of the galaxy and the gas fractions in the progenitors, where we find a dry merger at snapshot 122, just around the dramatic change of the galaxy shape. 

Galaxy shapes are predominantly triaxial between snapshot 65 and 105. This is because there are many wet mergers between these snapshots, which often make the galaxy shapes irregular. And the bars in these progenitors can make them appear triaxial, as shown in Section~\ref{sec:select_prolate}. In the bottom panel, we show the images, isophotes in three projections and line-of-sight velocity maps in one particular projection of the main progenitor of this galaxy at snapshot 115. It is oblate with strong rotation ($\sim$160 km/s).

The other prolate galaxy example is subhalo163932 (see Fig.~\ref{fig:subhalo163932}). It does not have clear rotation, and is slightly triaxial. Similarly, it used to be oblate between snapshots 95 to 115, but quickly becomes prolate after snapshot 125. In the merger tree, we can find a dry merger at snapshot 122. Its main progenitor is an oblate galaxy with strong rotation ($\sim$120 km/s).
 
\begin{figure*}
\includegraphics[width=\textwidth]{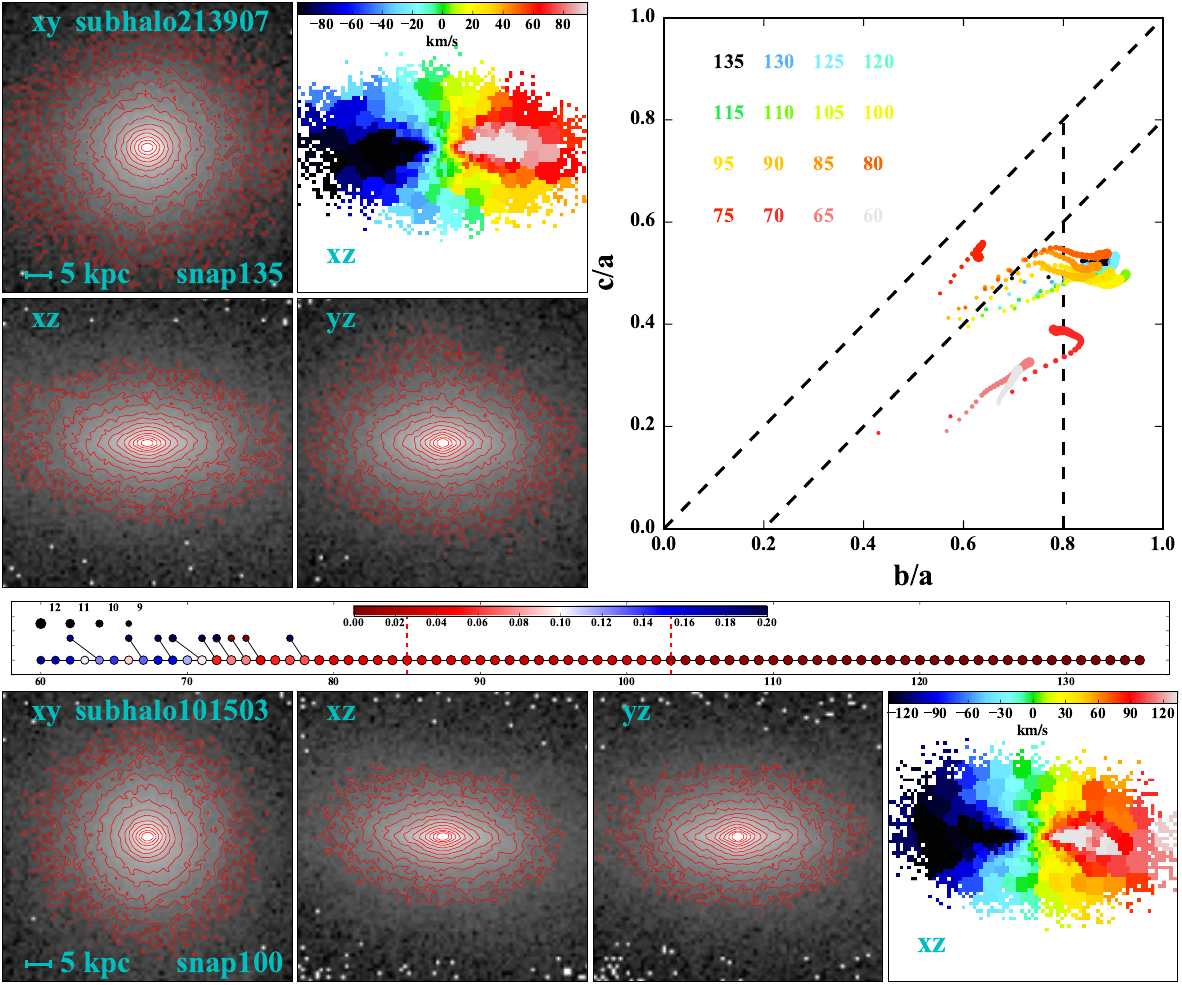}
    \caption{An example for an oblate galaxy, subhalo213907, with no dry merger. Labels
    are similar to Fig.~\ref{fig:subhalo210738}.}
    \label{fig:subhalo213907}
\end{figure*}

\begin{figure*}
\includegraphics[width=\textwidth]{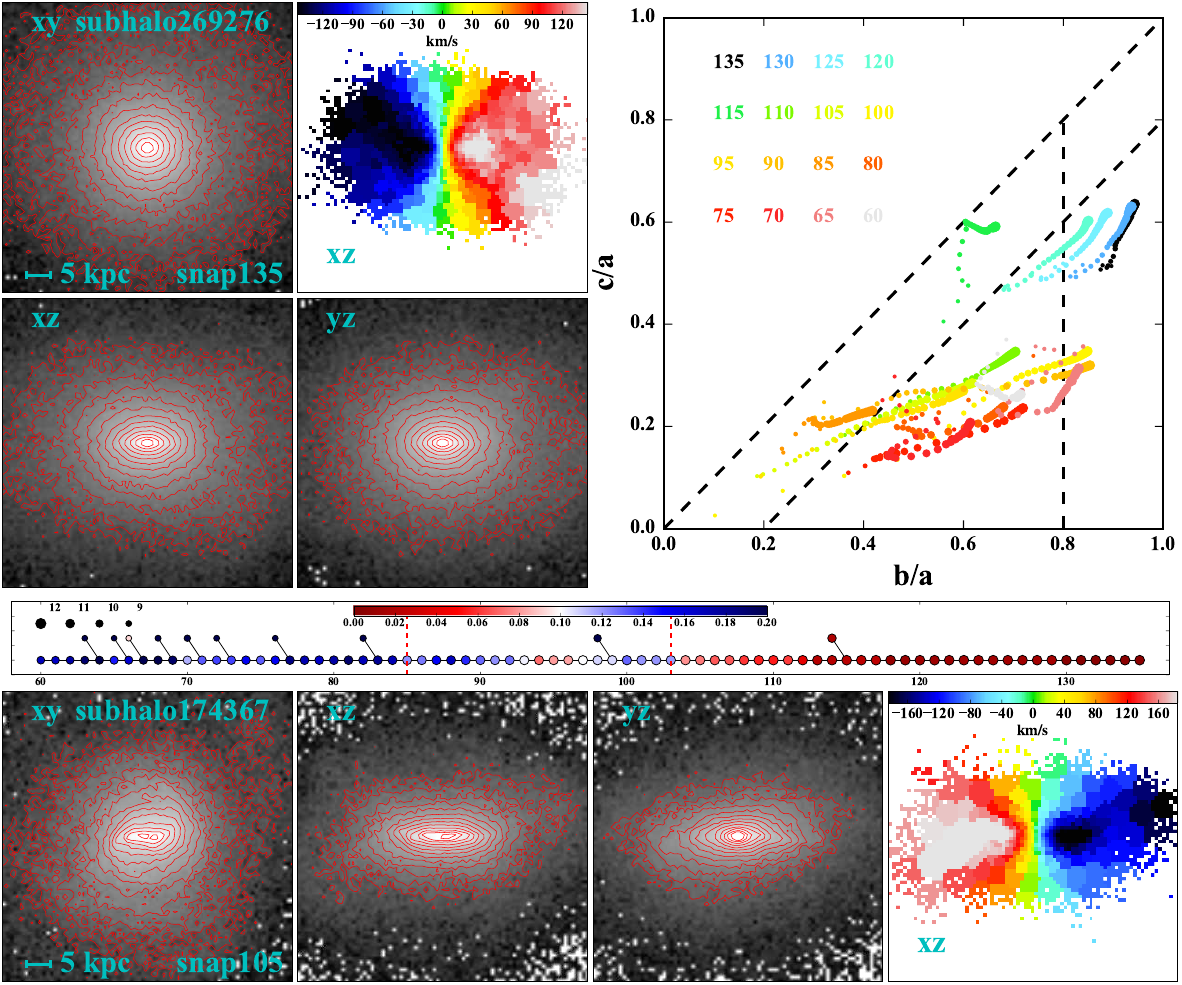}
    \caption{An example of an oblate galaxy, subhalo269276, with a dry merger
     at snapshot 115. Labels are similar to Fig.~\ref{fig:subhalo210738}.}
    \label{fig:subhalo269276}    
 \end{figure*}
 
We also visually examined the shape evolutions and the merger histories of all the prolate galaxies in our sample as well as some other galaxies with similar stellar mass. The images, velocity maps and merger trees for all the prolate galaxies are shown in Appendix \ref{AppendixA}. We find that all the prolate galaxies have at least one dry merger (mainly major, see Section~\ref{sec:mass_ratio} for mass ratios), except for one galaxy, subhalo277529, whose progenitor2 contains $\sim15\%$ gas, and progenitor1 contains less than $5\%$. The main progenitors (progenitor1) are usually fast rotating disk or oblate galaxies. After one or several dry mergers, they become a prolate galaxy with or without minor axis rotation. The other galaxies with similar stellar mass could also have had major dry mergers, but their shape are not prolate (see the text below for an example).

In Fig.~\ref{fig:subhalo213907} and Fig.~\ref{fig:subhalo269276}, we show similar features for two exemplary oblate galaxies. The first one is subhalo213907 (see Fig.~\ref{fig:subhalo213907}). It has no major dry merger in its evolution history. As can be seen, it remains of oblate shape until the end of the simulation. In another example, subhalo269276 is a disk galaxy at snapshot 105 (with bar or spiral-like structures). At snapshot 115, it has a dry merger, but the remnant is still a fast rotating oblate galaxy. We randomly examine 30 galaxies and find that galaxies with no major dry merger are all fast rotating oblate galaxies, like subhalo213907 in Fig.~\ref{fig:subhalo213907}. There are also some galaxies with one or more dry mergers, but the remnants are oblate, like subhalo269276 in  Fig.~\ref{fig:subhalo269276}.

\subsection{Stellar mass ratios in the dry mergers}
\label{sec:mass_ratio}
\begin{figure}
\includegraphics[width=\columnwidth]{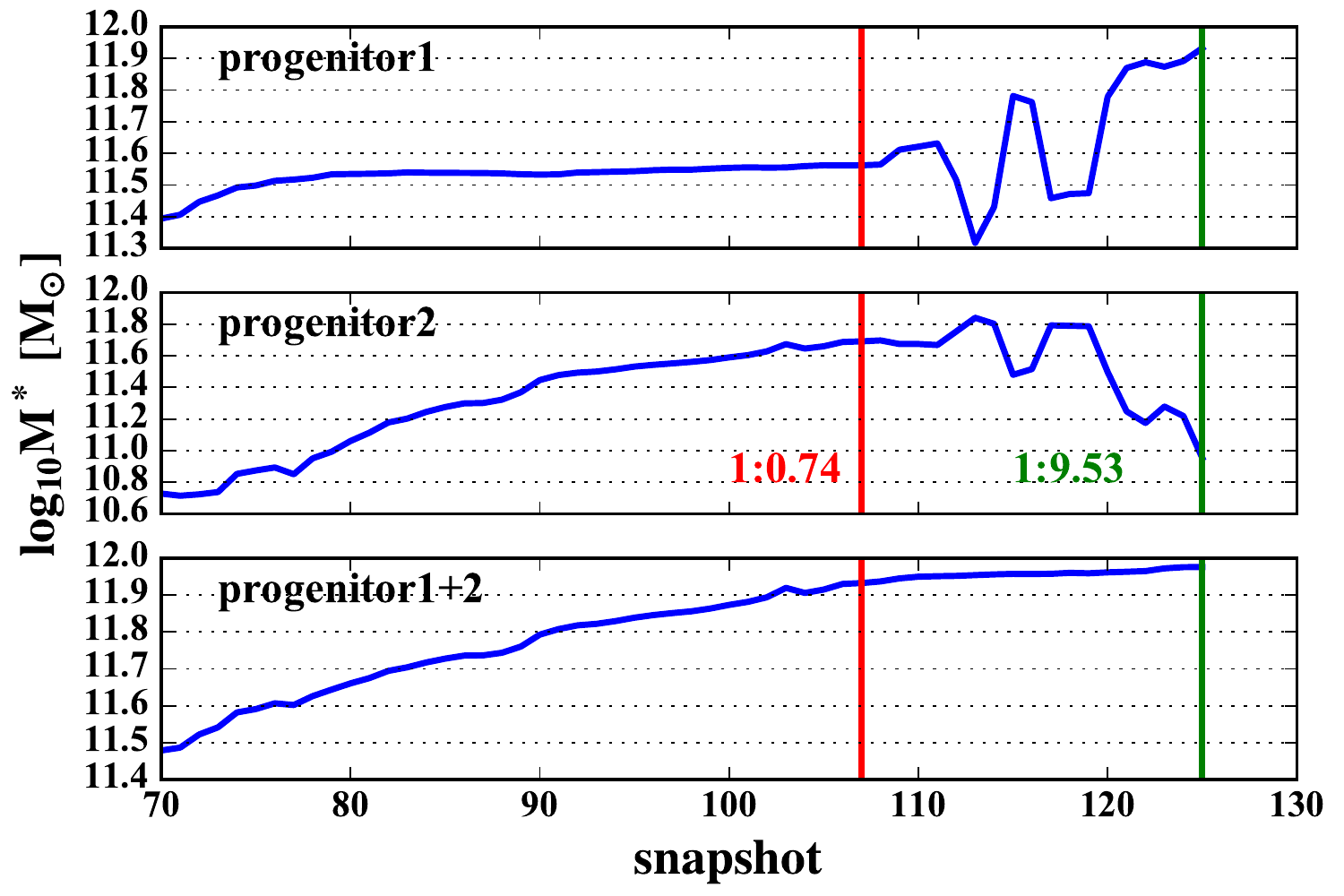}
    \caption{Example stellar mass growth profile (stellar mass vs. snapshot) for progenitor 1 (top), progenitor 2 (middle) and the sum of the two progenitors (bottom). The red and green vertical lines indicate, respectively, snapshot 107, after which the two stellar masses start oscillating wildly, and snapshot 125, where the two galaxies merge. The mass ratios at snapshot 107 (red) and 125 (green) are given in the middle panel.}
    \label{fig:measure_mratio}
\end{figure}

In this section, we check the stellar mass ratios in the dry mergers which produce oblate or prolate galaxies. However, it becomes difficult for the {\small SUBFIND} algorithm to properly separate the particles that belong to two progenitors when they get closer in distance. In Fig.~\ref{fig:measure_mratio}, we show the stellar mass growth curves (stellar masses at successive snapshots) of the progenitors of one galaxy that forms after a major merger event. As described in Section~\ref{sec:merger_history}, the more massive progenitor (from the immediate previous snapshot) is defined as progenitor1 (main progenitor), the less massive one is defined as progenitor2. The stellar mass growth curves of the two individual progenitors are shown in the top panel (progenitor1) and middle panel (progenitor2), respectively. In the bottom panel, we show the sum of the two progenitors. As can be seen, the stellar mass of progenitor1 stops growing after about snapshot 80. For progenitor2, it grows smoothly through minor mergers or in situ star formation. After snapshot 107 (red vertical lines), the individual stellar masses of the two progenitors oscillate significantly, until the two progenitors merge at snapshot 125 (green vertical lines). The summed stellar mass of the two progenitors, however, does not change significantly as can be seen in the bottom panel. This is because the merger is dry and there is nearly no star formation that would otherwise build up the stellar mass. It is worth noting that the mass ratio at snapshot 125 is 1:9.53 but 1:0.74 at snapshot 107 (for some galaxies like this one, the stellar mass of progenitor2 is even larger than progenitor1 at earlier snapshots due to the oscillation. In these cases, we just take the reciprocal of the mass ratios). 

If one just takes the stellar masses from the {\small SUBFIND} catalogue at one snapshot before the merger, the mass ratio would misleadingly indicate a minor merger event. In order to avoid such misidentifications and to accurately measure the mass ratios, we manually check the mass growth curves for all the galaxies of interest, and determine their stellar mass ratios one or two snapshots earlier before the \textit{oscillation period}. Since the mergers we check are mostly dry, there is nearly no star formation, and it is safe to trace back further in time to obtain the stellar mass ratios.

Another difficulty in measuring mass ratios is that multiple mergers can happen within a small time span. Once there is more than one merger within several snapshots, the definition of the mass ratio becomes ambiguous and it is much more difficult to obtain the correct mass ratio due to the oscillation problem described above. Below, we therefore only consider mass ratios of galaxies that have not suffered such multiple merger events in their histories. This results in 18 out of 35 prolate galaxies in our total sample.

\begin{figure}
\includegraphics[width=\columnwidth]{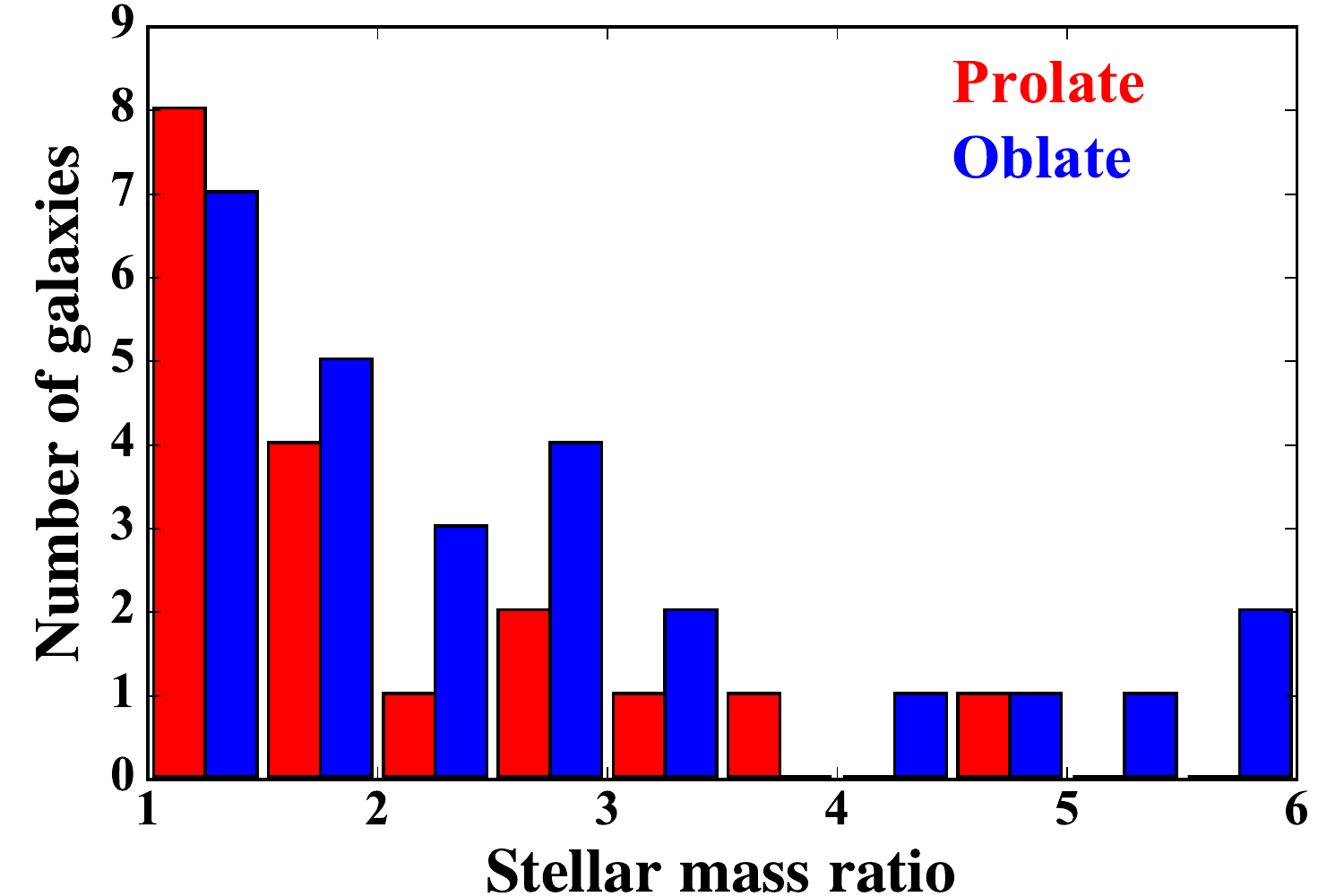}
    \caption{Stellar mass ratios of dry mergers for a selected prolate population (red) and for the oblate galaxies in the control sample (blue). Note that in order to have accurate measurements of mass ratios, both galaxy samples are selected not to have multiple merger events that happen within a small time span.}
    \label{fig:mratio}
\end{figure}

In addition to prolate galaxies, we select some oblate galaxies with dry mergers, which are also pruned from multiple merger events, as the control sample. In the selection, obvious minor mergers, i.e., based on the merger tree the mass ratio is smaller than $\sim 1/20$, are excluded since we would like to check whether major dry mergers can make oblate remnants, and we do not need a complete sample for this purpose. The stellar mass ratio distribution of the selected dry-merger prolates and oblates are shown in Fig.~\ref{fig:mratio}. As can be seen, 15 of a total of 18 prolate galaxies have mass ratios larger than 1:3, 2 are between 1:3 and 1:4, only one is between 1:4 and 1:5. While for oblate galaxies, both major and minor mergers exist. This shows that major dry mergers are responsible (but not exclusively) for the formation of prolate galaxies. 

\subsection{Merger orbit}
\label{sec:merger_orbit}
In this section, we discuss the merger orbits and try to understand the
differences between the dry mergers that produce oblate galaxies and prolate
galaxies. We treat the merging galaxies as point masses. We use the position of
the most bound particle as the position of a galaxy. The velocity of a galaxy
is calculated using the mass weighted mean velocities of the stellar particles
within the half stellar mass radius. We use the positions and velocities from
25 snapshots prior to the snapshot at which the two galaxies merge. We fail to
obtain a reliable direction of the orbital angular momentum (i.e. the direction
of the orbital plane), as we find many galaxies, especially prolate ones, do not have a well defined orbital plane. This is partly because when the two merging galaxies get close to each other, they will have other minor mergers, which may change the orbital angular momentum. In addition, other nearby halos also have effects on the merger orbits. Therefore, we choose to check the merger orbital type (radial or circular) instead of calculating the exact orbital angular momentum. 

\begin{figure}
\includegraphics[width=\columnwidth]{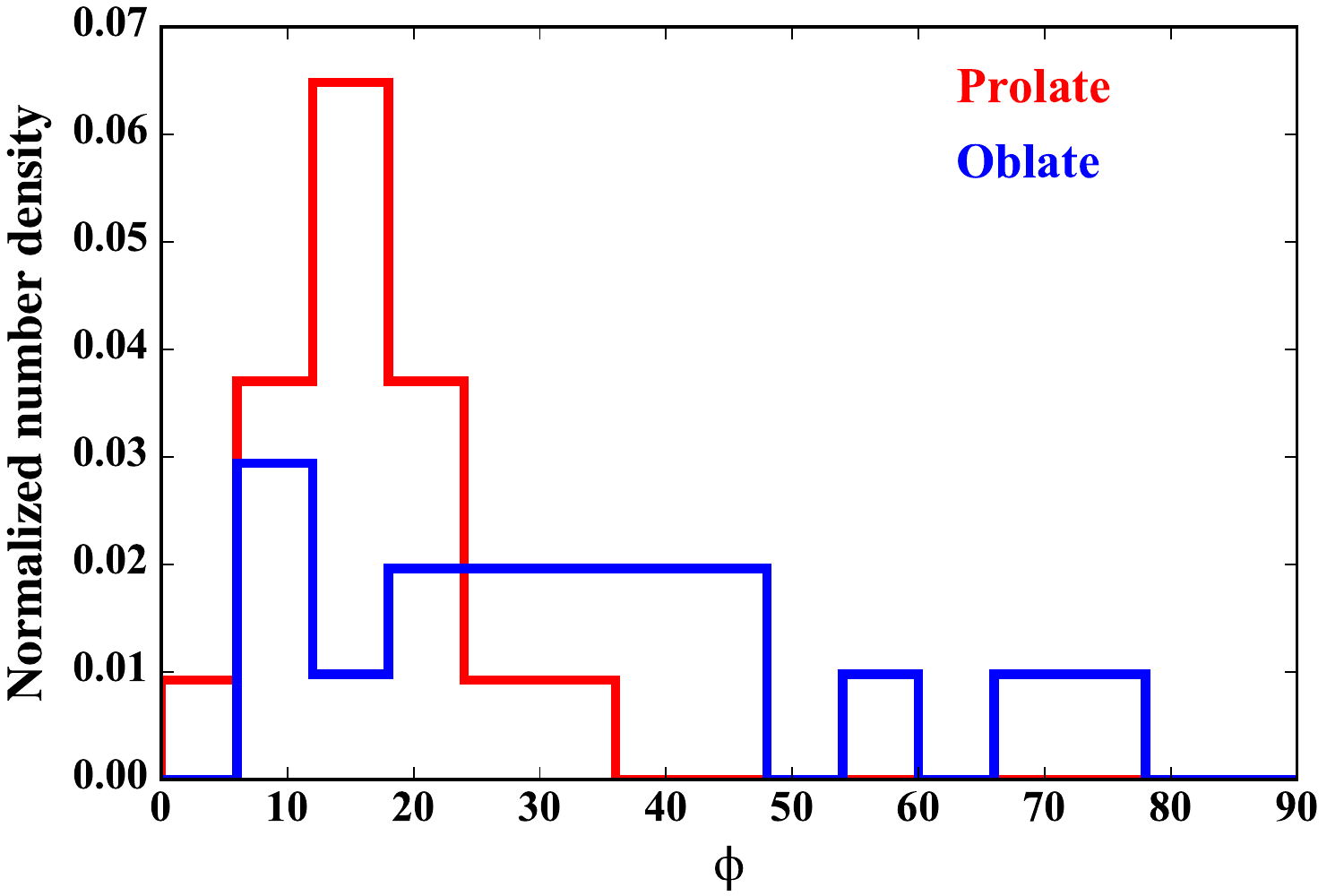}
\caption{Distribution of the angle $\phi$ for prolate (red) and oblate (blue) galaxies.  $\phi$ is defined as the mean angle between the relative velocity and the position of the two progenitors in the merger. If $\phi$ is close to 0, the merger orbit is radial. The prolate sample in this figure is the same as in Fig.~\ref{fig:mratio}, while the oblate sample is a subsample of those in Fig.~\ref{fig:mratio} where the merger mass ratios are required to be greater than $1:3$.}
    \label{fig:orbit}
\end{figure}

We calculate the angle between the relative velocity and the relative position of the two merging galaxies at every snapshot we choose. We define the angle $\phi$ as the median value of the angles at different snapshots. If $\phi$ is close to 0, the merger orbit is nearly radial, while a larger $\phi$ represents a more circular orbit. In Fig.~\ref{fig:orbit}, we plot the distribution of the angle $\phi$ for the prolate and oblate galaxies. The prolate sample is the same as in Fig.~\ref{fig:mratio}, while the oblate sample is a subsample of the oblate galaxies in Fig.~\ref{fig:mratio} where the merger mass ratios is required to be greater than $1:3$. This is to exclude the effects of minor mergers, which are unlikely to produce a prolate galaxy irrespective of their orbits. As can be seen, prolate galaxies usually have smaller angles (i.e. more radial merger orbits), while oblate galaxies can have nearly all different angles. Both radial and circular merger orbits can produce an oblate galaxy, however, the former tends to produce a slowly rotating system while the latter produces a rapidly rotating system (also see Section~\ref{sec:minor_axis_rotation} for more discussion). 

\citet{Ebrova2015} use N-body simulations of two identical disc galaxies to study the origin of the minor axis rotation in a dwarf spheroidal galaxy. They shows that the direction of the last encounter dominates the elongation of the remnant galaxy (their figure 3).  Due to the limitations of cosmological simulation, we do not have enough time resolution (i.e. enough snapshots) to catch the orbits around the last encounter. So here we are not able to make a direct comparison.

\subsection{Origin of the minor axis rotation}
\label{sec:minor_axis_rotation}
\begin{figure*}
\includegraphics[width=\textwidth]{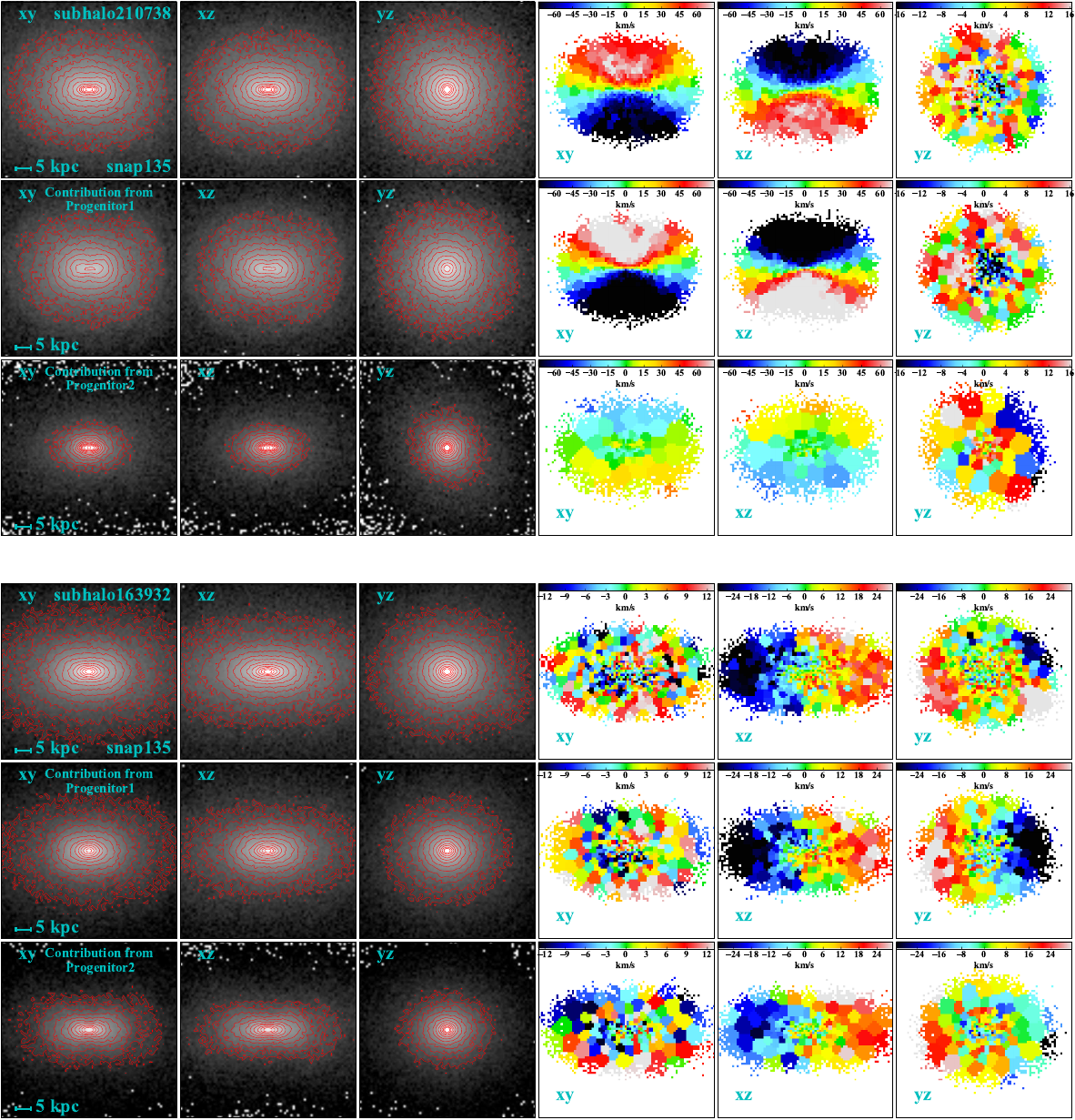}
    \caption{Two example prolate galaxies with major dry merger remnants are shown in the figure (Top panel: subhalo210738, with minor axis rotation. Bottom panel: subhalo163932, with no significant rotation). For each galaxy, the images and velocity maps along different projections of the whole galaxy (for snapshot 135, $z=0$) are shown in the first row. Those that are reconstructed by the particles that used to belong to progenitor1 and progenitor2 (from snapshot 115, $z=0.27$) are shown in the second and the third row, respectively. The colour scales are the same for images and velocity maps in the same column.}
    \label{fig:map_prolate}    
 \end{figure*}
 
 \begin{figure*}
\includegraphics[width=\textwidth]{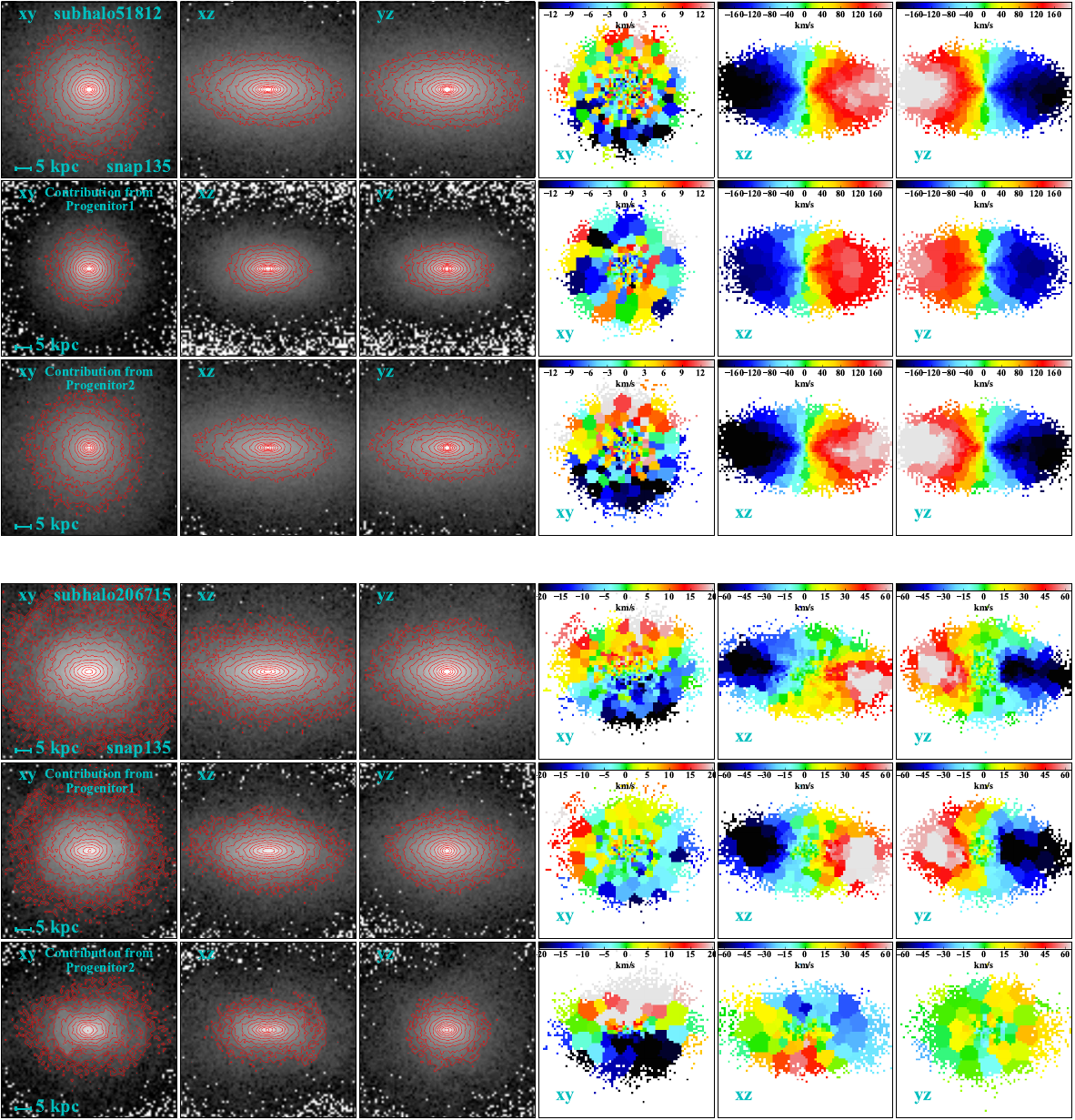}
\caption{Two example oblate galaxies as major dry merger remnants are shown in the figure (Top: subhalo51812, with a circular merger orbit. Bottom: subhalo206715, with a radial merger orbit). Pattern and labels are the same as in Fig.~\ref{fig:map_prolate}.}
    \label{fig:map_oblate}    
 \end{figure*}
 
As shown in Appendix \ref{AppendixA}, 18 out of 35 prolate galaxies in our sample have minor axis rotation (rotation is about the longest axis $a$), while the others have no significant rotation. The galaxies with minor-axis rotation are shown with bold font in Table~B1. In this section, we study the origin of such a minor-axis rotation in a prolate galaxy by examining the contribution of the rotation from its two progenitors. We select 9 prolate galaxies with clear minor axis rotation and simple merger history (i.e. no multiple major mergers within several snapshots) to do the study.
 
In the Illustris simulations, every particle has it unique ID throughout the whole simulation. For each of the nine prolate galaxies, we find out its progenitors at earlier snapshot (before the oscillation period described in Section~\ref{sec:mass_ratio} to avoid particle mixing) and take their particle IDs. By comparing the particle IDs between progenitors and the merger remnant (i.e. the prolate galaxy we are interested in), we can separate the particles of a prolate galaxy from its progenitor1 and progenitor2.  We then use the separated particles to reconstruct images and velocity maps using the same method as described in Section~\ref{sec:select_prolate}. Two examples are shown in Fig.~\ref{fig:map_prolate}, one has minor axis rotation and the other does not have clear rotation. The images and velocity maps along three different projections of the prolate galaxy (from snapshot 135, $z=0$) are shown in the first row. Those that are reconstructed using the separated particles that used to belong to progenitor1 and progenitor2 (from snapshot 115, $z=0.27$) are shown in the second and the third row, respectively.
 
As can be seen in the figure, for prolate galaxy subhalo210738 (top panels), the rotation velocities are $\sim 80\, {\rm km\,s^{-1}}$ and mainly dominated by the contribution from progenitor1. Progenitor2 contributes little in the central part, and in the outer part the rotation direction is opposite to the rotation in progenitor1, although it is quite noisy. If the angular momentum of the minor-axis rotation (of the prolate remnant) would have resulted from the orbital angular momentum during the merger, the angular momenta of the two constituents should have similar patterns in the remnant. However, this is not the case. In addition, from the images, one can see that the particles from progenitor1 are more extended, while the particles from progenitor2 are more centrally concentrated, representing the core region of the whole galaxy. Note that the majority of the nine prolate galaxies with clear minor-axis rotation are similar to subhalo210738, where the rotation is clearly dominated by particles from one of the two progenitors (and usually it is the more massive progenitor, i.e. progenitor1). Prolate galaxy subhalo163932 (bottom panels) does not have minor axis rotation, and the angular momenta of its oblate progenitor1 and progenitor2 have mostly been lost during the merger.
 
 \begin{table}
\centering
\label{cosi_vect}
\tabcolsep=15.0pt
\caption{The angle between the spin vector of the minor-axis rotation prolate galaxies and the spin vector of the oblate progenitor which contributes to the prolate rotation in a dominant manner.}
\begin{tabular}{@{}lclc@{}}
  \hline
  \hline
  galaxy ID & $\cos i$ & galaxy ID & $\cos i$ \\
  \hline
  subhalo123773& 0.447& subhalo129771 & 0.893\\
  subhalo138413& 0.134& subhalo152864 & 0.953\\
  subhalo185229& 0.843& subhalo210738 & 0.982\\
  subhalo222715& 0.927& subhalo225517 & 0.901\\
  subhalo277529& 0.862\\
  \hline
\end{tabular}
\begin{minipage}{8.4cm}
  Notes:
  For subhalo222715, it has less rotation than the other galaxies in the table. This
  is because this galaxy includes two counter rotational components, which come from each of 
  the two progenitors, with opposite spin directions.
\end{minipage}
\end{table}

For each of the nine prolate galaxies with clear minor-axis rotation and simple merger history, we calculate the angle $i$ between the spin vector of the prolate galaxy and the spin vector of the oblate progenitor which dominates the prolate rotation in a dominant manner. The spin vector is calculated with the stellar particles within the half stellar mass radius. The centre of the galaxy is chosen as the position of the most bound particle in that galaxy. The $\cos i$ values of these angles are listed in Table~1. As one can see, subhalo123773 and subhalo138413 have smaller $\cos i$. We believe only part of the spin angular momenta of their progenitors is converted into the final minor axis rotation. For most galaxies, $\cos i$ is close to 1 (i.e. the two spin vectors have similar direction). This suggests that most of the minor axis rotation comes from the spin angular momentum of their dominant oblate progenitors. 

\citet{Ebrova2015} found that the angular momentum of the minor axis rotation comes from the spin angular momentum of the progenitors, and a near radial merger orbit is required to produce strong minor axis rotation. This is consistent with our results, which are in a cosmological context with more realistic merger initial conditions, environments and physics.

In comparison, Fig.~\ref{fig:map_oblate} shows the reconstructed images and velocity maps of two example oblate galaxies as major merger remnants. The first one is subhalo51812 (top panels), which has a mean rotation velocity of $\sim 200\, \rm km/s$ and is formed via a circular merger orbit. The second one is subhalo206715, which has a mean rotation velocity of $\sim 70\, \rm km/s$ and is formed via a radial merger orbit. As shown in Fig.~\ref{fig:orbit}, both radial and circular orbits can produce an oblate galaxy. However, as shown in the figure, oblates that are formed via circular merger orbits tend to rotate much faster, and the two progenitors end up with similar contributions to this final rotation. In contrast, oblate galaxies that are formed via radial merger orbits tend to rotate slower; the contribution of the rotation mainly comes from progenitor1, while the contribution from progenitor2 even has a little minor axis rotation.

In addition to the minor-axis rotation, there are also some other interesting features in the prolate sample, e.g. oblate rotation (rotation is about the minor axis) in subhalo73663, subhlao163932, subhalo200653 and subhalo217716, counter rotation in subhalo129770, subhalo138413 and subhalo222715, and kinematically decoupled cores in subhalo129771, subhalo165890 and subhalo177128. We note that by a similar analysis of reconstructed  velocity maps, we find that the kinematically decoupled cores are contributed by the particles from one of the progenitors, instead of from two different progenitors.

We mention in passing that all the findings above have their roots in the detailed configurations of the spins of the incoming merger progenitors, the merger orbital angular momentum, and the spin of the final merger remnants. However, the output spacing of the Illustris simulation is not short enough to reliably make time-resolved measurements. We therefore leave this issue to a future work.

\subsection{Properties of prolate galaxies}
\label{sec:properties}

\begin{figure*}
\includegraphics[width=\textwidth]{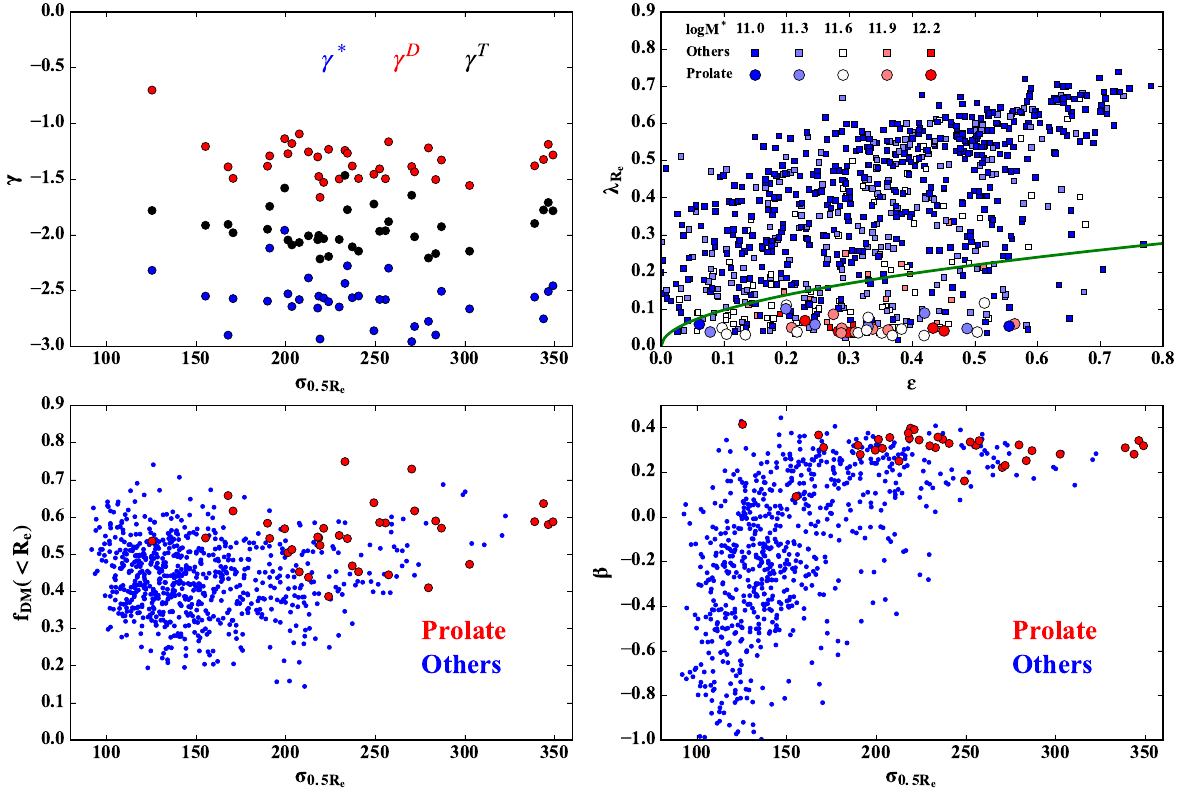}
    \caption{Upper left: inner density slopes vs. velocity dispersion for the 35 prolate galaxies in the sample. Blue, red and black represent the stellar, the dark matter and the total inner density slopes, respectively. Upper right: parameter $\lambda_{R_e}$ vs. ellipticity. Big circles represent the 35 prolate galaxies, small squares represent the other 804 galaxies. Different colours represent different stellar masses, as labelled within the panel. The solid green line is 0.31$\times \sqrt{\epsilon}$, which is used to separate fast rotators and slow rotators in \citet{Emsellem2011}. Lower panels: dark matter fraction within $R_e$  vs. velocity dispersion (left-hand side) and stellar orbital anisotropy parameter $\beta$ vs. velocity dispersion (right-hand side). In both panels, red and blue represent the 35 prolate galaxies and the other 804 galaxies in the sample, respectively. }
    \label{fig:prop}
\end{figure*}

We investigate some important properties and relations of the prolate galaxies in our sample, focussing on velocity dispersion, effective radius, average ellipticity, $\lambda_{\rm R}$ \citep{Emsellem2007}, spherical anisotropy parameter $\beta$ \citep{Binney2008}, the dark matter fraction within $R_e$, and the inner stellar, dark matter and total mass density slopes. We choose the $z$-direction in the simulation coordinate system as line-of-sight direction for all the galaxies. This ensures our projection direction is random relative to the orientation of the galaxies. 

To calculate the effective radius and ellipticity, we first create mock images from stellar mass maps, assuming the stellar mass-to-light ratio equals  1 (the method is described in Section~\ref{sec:select_prolate}, with 0.5 kpc/pixel grid resolution). We define the effective radius as 
\begin{equation}
R_e = \sqrt{A_e/\pi},
\end{equation}
where $A_e$ is the area of the isophote which contains half of the total luminosity (i.e. stellar mass). We note that uncertainties in $R_e$ have insignificant effects on measurements of $\epsilon$ and $\lambda_{\rm R}$, as the latter does not sensitively depend on the boundary, as defined by the former, within which the measurements are made.

Similar to \citet{Cappellari2007}, we define the ellipticity as:
\begin{equation}
\epsilon = 1-\sqrt{\frac{\sum_{1}^{N} f_n y_n^2}{\sum_{1}^{N} f_n y_n^2}},
\end{equation}
where $f_n$ is the flux (i.e. stellar mass) within the $n$-th image pixel, and $x_n$ and $y_n$ are the coordinates of the $n$-th image pixel. The coordinates are centred on the position of the galaxy (defined as the minimum of its gravitational potential), and the $x$ and $y$-axes are aligned with the major and  minor projection axes, respectively. The sum is over all the pixels within the isophote which includes half of the total luminosity.

We follow the practice of \citet{Emsellem2007} to calculate the parameter $\lambda_{\rm R}$. We first create Voronoi-binned velocity and velocity dispersion maps for each galaxy using the same method described in Section~\ref{sec:select_prolate}. The parameter $\lambda_{\rm R}$ is then calculated using the Voronoi-binned velocity maps within the ellipse that has ellipticity $\epsilon$ and that encloses half of the total luminosity. 

The anisotropy parameter $\beta$ is measured for stellar particles within a 3D radius of the effective radius, available from \citet{Xu2017} (see their Eqn.~14 for the precise definition). The dark matter fraction within $R_e$ is defined as the dark matter mass within a 3D radius of $R_e$ from the galaxy centre divided by the total mass within the same radial range. For the density slopes, we use the galaxy particle data to calculate the radial mass density profiles $\rho(r)$ for stellar, dark matter and total mass. We then use a linear function to fit $\log r$ and $\log \rho$ between $0.1r^{*}_{\rm h}$ and $0.5r^{*}_{\rm h}$. The logarithmic density slope $\gamma$ is defined as the best fitting slope of the linear function. We note that these inner density slopes are quantitatively consistent with those from \citet{Xu2017}, albeit the investigated radial ranges are slightly different. All the calculated properties are listed in table~B1.

In Fig.~\ref{fig:prop}, we show the velocity-dispersion dependencies of the inner density slopes (upper left), dark matter fractions within $R_e$ (lower left), anisotropy parameter $\beta$ (lower right), as well as the $\epsilon-\lambda_{R_e}$ relation (upper right). The slopes do not seem to correlate well with the velocity dispersion of the host galaxy for the prolate galaxies. The total density slopes are close to isothermal (i.e. $\gamma\sim-2.0$), which is similar to the results in \citet{Remus2017}. The stellar (dark matter) density slopes are steeper (shallower). For these prolate galaxies, the dark matter density slopes are steeper than the standard NFW \citep{Navarro1996} prediction \citep[see also][]{Xu2017}. 

The dark matter fractions of prolate galaxies do not show a significant difference compared with other galaxies in the sample. Correspondingly, all the prolate galaxies have radial stellar orbital anisotropies ($\beta > 0$), which echo their radial merger orbits emphasised in this work. This
is consistent with the orbital structure of collisionless merger remnants in \citet{Jesseit2005}
and cosmological zoom simulation in \citet{Rottgers2014}.
We also check the other galaxies with similarly large $\beta$-values, and find that many of them have strong bars, which contain more radial orbits than the disk component.  As shown in the $\epsilon-\lambda_{R_e}$ diagram, our prolate galaxies are all slow rotators as defined in \citet{Emsellem2011}. The galaxy distribution in the diagram, however, is not similar to the results in observations, e.g. $\rm ATLAS^{3D}$ \citep{Emsellem2011}, CALIFA \citep{Falcon2015},
SAMI \citep{Fogarty2015}, MASSIVE \citep{Veale2017b,Veale2017a} and 
MaNGA (Graham 2017, in preparation).
All these observations have very few slow rotators with $\epsilon>0.4$, while the prolate galaxies in
the Illustris simulation could have $\epsilon$ as large as $0.55$.
A similar problem has been pointed out in \citet{Naab2003}, which found collisionless binary mergers 
with equal mass produce remnants with ellipticity higher than those seen in observations. 
The higher ellipticity in the simulation may be because the merger is too dry \citep[class E]{Naab2014}
or the resolution is not higher enough \citep{Bois2010,Bois2011}. \citet{Moody2014} also pointed out 
that multiple major mergers could produce less elongated galaxies than binary mergers.

\section{Conclusions}
\label{sec:conclusions}
We study galaxy shapes in the Illustris cosmological hydrodynamic simulation. The galaxy sample we use is selected by stellar mass ($\rm M^* > 10^{11} M_{\odot}$) and S{\`e}rsic index ($n_{\rm S{\`e}rsic} > 2.0$), yielding a total of 839 galaxies.  We use the reduced inertial mass tenser method to measure a galaxy's axis ratios, combined with visual checks, and divide the galaxies into oblate, triaxial and prolate galaxies. We find that massive galaxies tend to be prolate, and there is nearly no prolate galaxy with stellar mass smaller than $3\times10^{11}\rm M_{\odot}$ in our sample.

We use merger trees extracted from the simulation to examine the formation history of those massive prolate galaxies, and find that nearly all the massive prolate galaxies have major dry merger in the past. The mass ratios are usually between $1:1$ and $1:3$. The main progenitors are usually disks or fast rotating oblate galaxies before the merger, and become prolate soon after. In addition, we check the merger orbits of these prolate galaxies as well as some oblate galaxies which have major dry merger in their past. We find that most prolate galaxies had a more radial merger orbit (angle $\phi<20^{\circ}$, see Section~\ref{sec:merger_orbit} for the definition), while oblate galaxies could have had either radial or circular merger orbits. A few of the prolate galaxies produced by major dry mergers have higher ellipticity than those in observations ($\epsilon<0.4$ for most slow rotators
as observed by different surveys). This may be because the merger is too dry \citep{Naab2014}, the 
resolution is not high enough and/or multiple major mergers instead of a single merger are required
\citep{Moody2014}. 

\citet{Rodriguez-Gomez2015} find that the long axis of a prolate galaxy is aligned with the direction of the last encounter of its merger orbit using N-body simulations (see their figure 3). Cosmological simulations, however, do usually not have enough output times to accurately measure that direction and compare with their results directly. This could be investigated in our future works using similar outputting frequency as in their simulations. In addition, studies of galaxy orientations and their large scale environments (e.g. \citealt{Zhang2013,Shi2015,Welker2014}) find that the major axes of galaxies in filaments tend to be aligned with the directions of the filaments, and galaxies in sheets have their major axes parallel to the plane of the sheets. And the alignments are stronger for red central galaxies. These are consistent with the formation mechanism of the prolate galaxies and the mass dependence that we find.

Some prolate galaxies show clear minor axis rotation. In order to understand the origin of such rotation, we find the particles of a prolate galaxy from different progenitors. We then check their contribution to the minor axis rotation. We find that the angular momentum of such rotation usually comes from the spin angular momentum of the progenitors (usually main progenitor). This is consistent with the results of N-body simulations in \citet{Rodriguez-Gomez2015}.

\citet{Penoyre2017} studied the origin of slow and fast rotators in the Illustris simulation, but 
they did not specifically focus on prolate galaxies. And they did not study the effects of merger
orbits on the remnant properties. In their work, they found major merger is the main cause of the 
slow rotators, while in some rare cases the remnants are also spun up. In our work, we pointed
out that the rotation properties are related to the merger orbital parameters. A major merger with a circular 
orbit could produce a fast rotator. In addition, due to the modest sample in our study 
(35 prolate galaxies), we were able to study more detailed relations between merger history
and galaxy properties (e.g. shape and rotation) by visual examination.

The mass resolution of the Illustris simulation is $\sim 10^6\, {\rm M}_{\odot}$. The galaxies we selected all have stellar masses larger than $10^{11}\, {\rm M}_{\odot}$, so that there are more than $10^5$ stellar particles in each galaxy. This ensures that the galaxies we study have 
enough particles to accurately measure their shape at different radii.
Also note that the softening length of the simulation is 710 pc, which is much smaller than the galaxies we study. \citet{Bois2010,Bois2011} pointed out that the numerical resolution 
could have significant effects on the shape of the remnant in a gas-rich merger, with higher 
resolution producing rounder system \citep[see also][Class C]{Naab2014}.
For dry mergers, however, the effects are visible,
but not significant. Thus we expect that the dry mergers which produce the prolate galaxies in the
simulation are reliable with respect to the resolution, while the gas rich processes at higher
redshift could be affected, which might slightly change the properties of those prolate galaxies
at redshift 0 (e.g. ellipticity).
The feedback model in the simulation, however, may have effects on the gas fraction evolution, 
baryon conversion efficiencies \citep[figure 11]{Pillepich2017},
as well as the number of the galaxies with low gas fractions \citep{Martizzi2012, Dubois2016}.
A few prolate galaxies with high ellipticity not seen in observations could be due
to such low gas fractions.
Improved versions of the Illustris simulation model \citep{Pillepich2017, Weinberger2017} will use updated feedback models and larger box size than the current simulation. In future works, we could compare the evolution history and the formation mechanism of the prolate galaxies between the present model and these forthcoming simulations. Larger galaxy samples in future simulations would also be extremely useful to have better statistics for studying the dependence of the merger orbits and galaxy orientations on the large-scale environments (e.g. clusters, filaments, sheets or voids).

Observationally, it will be interesting to examine the properties of these massive galaxies like dark matter fraction within the effective radius \citep{Cappellari2013b}, the position on the fundamental plane (Li et al., in preparation), the age, metallicity gradient and their large scale environment \citep{Zheng2017} and the stellar initial mass function \citep{Cappellari2012,Li2017}. Theoretically, it is useful to understand the exact orbital parameters of major dry mergers that produce a prolate galaxy, the correlation of these parameters with the properties of the merger remnants, and the probability of forming a prolate galaxy. This can be done in future works with N-body simulations similar to \citet{Rodriguez-Gomez2015}.

\section*{Acknowledgements}

HL would like to thank Dr. Dylan Nelson for advice on the Illustris data, Michele Cappellari and the anonymous referee for comments that improved the paper. We performed our computer runs on the Zen high performance computer cluster of the National Astronomical Observatories, Chinese Academy of Sciences (NAOC). This work was supported by the National Science Foundation of China (Grant No. 11333003, 11390372 to SM). VS acknowledges support through the European Research Council under ERCStG grant EXAGAL-308037. DDX and VS would like to thank the Klaus Tschira Foundation.
 



\bibliographystyle{mnras}
\bibliography{lhy} 

\begin{thebibliography}{}
\makeatletter
\relax
\def\mn@urlcharsother{\let\do\@makeother \do\$\do\&\do\#\do\^\do\_\do\%\do\~}
\def\mn@doi{\begingroup\mn@urlcharsother \@ifnextchar [ {\mn@doi@}
  {\mn@doi@[]}}
\def\mn@doi@[#1]#2{\def\@tempa{#1}\ifx\@tempa\@empty \href
  {http://dx.doi.org/#2} {doi:#2}\else \href {http://dx.doi.org/#2} {#1}\fi
  \endgroup}
\def\mn@eprint#1#2{\mn@eprint@#1:#2::\@nil}
\def\mn@eprint@arXiv#1{\href {http://arxiv.org/abs/#1} {{\tt arXiv:#1}}}
\def\mn@eprint@dblp#1{\href {http://dblp.uni-trier.de/rec/bibtex/#1.xml}
  {dblp:#1}}
\def\mn@eprint@#1:#2:#3:#4\@nil{\def\@tempa {#1}\def\@tempb {#2}\def\@tempc
  {#3}\ifx \@tempc \@empty \let \@tempc \@tempb \let \@tempb \@tempa \fi \ifx
  \@tempb \@empty \def\@tempb {arXiv}\fi \@ifundefined
  {mn@eprint@\@tempb}{\@tempb:\@tempc}{\expandafter \expandafter \csname
  mn@eprint@\@tempb\endcsname \expandafter{\@tempc}}}

\bibitem[\protect\citeauthoryear{{Allgood}, {Flores}, {Primack}, {Kravtsov},
  {Wechsler}, {Faltenbacher}  \& {Bullock}}{{Allgood}
  et~al.}{2006}]{Allgood2006}
{Allgood} B.,  {Flores} R.~A.,  {Primack} J.~R.,  {Kravtsov} A.~V.,  {Wechsler}
  R.~H.,  {Faltenbacher} A.,   {Bullock} J.~S.,  2006, \mn@doi [\mnras]
  {10.1111/j.1365-2966.2006.10094.x}, \href
  {http://adsabs.harvard.edu/abs/2006MNRAS.367.1781A} {367, 1781}

\bibitem[\protect\citeauthoryear{{Binney} \& {Tremaine}}{{Binney} \&
  {Tremaine}}{2008}]{Binney2008}
{Binney} J.,  {Tremaine} S.,  2008, {Galactic Dynamics: Second Edition}.
Princeton University Press

\bibitem[\protect\citeauthoryear{{Bois} et~al.,}{{Bois}
  et~al.}{2010}]{Bois2010}
{Bois} M.,  et~al., 2010, \mn@doi [\mnras] {10.1111/j.1365-2966.2010.16885.x},
  \href {http://adsabs.harvard.edu/abs/2010MNRAS.406.2405B} {406, 2405}

\bibitem[\protect\citeauthoryear{{Bois} et~al.,}{{Bois}
  et~al.}{2011}]{Bois2011}
{Bois} M.,  et~al., 2011, \mn@doi [\mnras] {10.1111/j.1365-2966.2011.19113.x},
  \href {http://adsabs.harvard.edu/abs/2011MNRAS.416.1654B} {416, 1654}

\bibitem[\protect\citeauthoryear{{Bryant} et~al.,}{{Bryant}
  et~al.}{2015}]{Bryant2015}
{Bryant} J.~J.,  et~al., 2015, \mn@doi [\mnras] {10.1093/mnras/stu2635}, \href
  {http://adsabs.harvard.edu/abs/2015MNRAS.447.2857B} {447, 2857}

\bibitem[\protect\citeauthoryear{{Bundy} et~al.,}{{Bundy}
  et~al.}{2015}]{Bundy2015}
{Bundy} K.,  et~al., 2015, \mn@doi [\apj] {10.1088/0004-637X/798/1/7}, \href
  {http://adsabs.harvard.edu/abs/2015ApJ...798....7B} {798, 7}

\bibitem[\protect\citeauthoryear{{Cappellari}}{{Cappellari}}{2016}]{Cappellari2016}
{Cappellari} M.,  2016, \mn@doi [\araa] {10.1146/annurev-astro-082214-122432},
  \href {http://adsabs.harvard.edu/abs/2016ARA%26A..54..597C} {54, 597}

\bibitem[\protect\citeauthoryear{{Cappellari} \& {Copin}}{{Cappellari} \&
  {Copin}}{2003}]{Cappellari2003}
{Cappellari} M.,  {Copin} Y.,  2003, \mn@doi [\mnras]
  {10.1046/j.1365-8711.2003.06541.x}, \href
  {http://adsabs.harvard.edu/abs/2003MNRAS.342..345C} {342, 345}

\bibitem[\protect\citeauthoryear{{Cappellari} et~al.,}{{Cappellari}
  et~al.}{2007}]{Cappellari2007}
{Cappellari} M.,  et~al., 2007, \mn@doi [\mnras]
  {10.1111/j.1365-2966.2007.11963.x}, \href
  {http://adsabs.harvard.edu/abs/2007MNRAS.379..418C} {379, 418}

\bibitem[\protect\citeauthoryear{{Cappellari} et~al.,}{{Cappellari}
  et~al.}{2011}]{Cappellari2011}
{Cappellari} M.,  et~al., 2011, \mn@doi [\mnras]
  {10.1111/j.1365-2966.2010.18174.x}, \href
  {http://adsabs.harvard.edu/abs/2011MNRAS.413..813C} {413, 813}

\bibitem[\protect\citeauthoryear{{Cappellari} et~al.,}{{Cappellari}
  et~al.}{2012}]{Cappellari2012}
{Cappellari} M.,  et~al., 2012, \mn@doi [\nat] {10.1038/nature10972}, \href
  {http://adsabs.harvard.edu/abs/2012Natur.484..485C} {484, 485}

\bibitem[\protect\citeauthoryear{{Cappellari} et~al.,}{{Cappellari}
  et~al.}{2013}]{Cappellari2013b}
{Cappellari} M.,  et~al., 2013, \mn@doi [\mnras] {10.1093/mnras/stt562}, \href
  {http://adsabs.harvard.edu/abs/2013MNRAS.432.1709C} {432, 1709}

\bibitem[\protect\citeauthoryear{{Carter}, {Thomson}  \& {Hau}}{{Carter}
  et~al.}{1998}]{Carter1998}
{Carter} D.,  {Thomson} R.~C.,   {Hau} G.~K.~T.,  1998, \mn@doi [\mnras]
  {10.1046/j.1365-8711.1998.01287.x}, \href
  {http://adsabs.harvard.edu/abs/1998MNRAS.294..182C} {294, 182}

\bibitem[\protect\citeauthoryear{{Dubois}, {Peirani}, {Pichon}, {Devriendt},
  {Gavazzi}, {Welker}  \& {Volonteri}}{{Dubois} et~al.}{2016}]{Dubois2016}
{Dubois} Y.,  {Peirani} S.,  {Pichon} C.,  {Devriendt} J.,  {Gavazzi} R.,
  {Welker} C.,   {Volonteri} M.,  2016, \mn@doi [\mnras]
  {10.1093/mnras/stw2265}, \href
  {http://adsabs.harvard.edu/abs/2016MNRAS.463.3948D} {463, 3948}

\bibitem[\protect\citeauthoryear{{Ebrov{\'a}} \& {{\L}okas}}{{Ebrov{\'a}} \&
  {{\L}okas}}{2015}]{Ebrova2015}
{Ebrov{\'a}} I.,  {{\L}okas} E.~L.,  2015, \mn@doi [\apj]
  {10.1088/0004-637X/813/1/10}, \href
  {http://adsabs.harvard.edu/abs/2015ApJ...813...10E} {813, 10}

\bibitem[\protect\citeauthoryear{{Emsellem}, {Monnet}  \& {Bacon}}{{Emsellem}
  et~al.}{1994}]{Emsellem1994}
{Emsellem} E.,  {Monnet} G.,   {Bacon} R.,  1994, \aap, \href
  {http://adsabs.harvard.edu/abs/1994A%26A...285..723E} {285, 723}

\bibitem[\protect\citeauthoryear{{Emsellem} et~al.,}{{Emsellem}
  et~al.}{2007}]{Emsellem2007}
{Emsellem} E.,  et~al., 2007, \mn@doi [\mnras]
  {10.1111/j.1365-2966.2007.11752.x}, \href
  {http://adsabs.harvard.edu/abs/2007MNRAS.379..401E} {379, 401}

\bibitem[\protect\citeauthoryear{{Emsellem} et~al.,}{{Emsellem}
  et~al.}{2011}]{Emsellem2011}
{Emsellem} E.,  et~al., 2011, \mn@doi [\mnras]
  {10.1111/j.1365-2966.2011.18496.x}, \href
  {http://adsabs.harvard.edu/abs/2011MNRAS.414..888E} {414, 888}

\bibitem[\protect\citeauthoryear{{Emsellem}, {Krajnovi{\'c}}  \&
  {Sarzi}}{{Emsellem} et~al.}{2014}]{Emsellem2014}
{Emsellem} E.,  {Krajnovi{\'c}} D.,   {Sarzi} M.,  2014, \mn@doi [\mnras]
  {10.1093/mnrasl/slu140}, \href
  {http://adsabs.harvard.edu/abs/2014MNRAS.445L..79E} {445, L79}

\bibitem[\protect\citeauthoryear{{Falc{\'o}n-Barroso}, {Lyubenova}  \& {van de
  Ven}}{{Falc{\'o}n-Barroso} et~al.}{2015}]{Falcon2015}
{Falc{\'o}n-Barroso} J.,  {Lyubenova} M.,   {van de Ven} G.,  2015, in
  {Cappellari} M.,  {Courteau} S.,  eds,  IAU Symposium Vol. 311, Galaxy Masses
  as Constraints of Formation Models. pp 78--81 (\mn@eprint {arXiv}
  {1409.7786}), \mn@doi{10.1017/S1743921315003439}

\bibitem[\protect\citeauthoryear{{Fogarty} et~al.,}{{Fogarty}
  et~al.}{2015}]{Fogarty2015}
{Fogarty} L.~M.~R.,  et~al., 2015, \mn@doi [\mnras] {10.1093/mnras/stv2060},
  \href {http://adsabs.harvard.edu/abs/2015MNRAS.454.2050F} {454, 2050}

\bibitem[\protect\citeauthoryear{{Franx}}{{Franx}}{1988}]{Franx1988}
{Franx} M.,  1988, \mn@doi [\mnras] {10.1093/mnras/231.2.285}, \href
  {http://adsabs.harvard.edu/abs/1988MNRAS.231..285F} {231, 285}

\bibitem[\protect\citeauthoryear{{Genel} et~al.,}{{Genel}
  et~al.}{2014}]{Genel2014}
{Genel} S.,  et~al., 2014, \mn@doi [\mnras] {10.1093/mnras/stu1654}, \href
  {http://adsabs.harvard.edu/abs/2014MNRAS.445..175G} {445, 175}

\bibitem[\protect\citeauthoryear{{Illingworth}}{{Illingworth}}{1977}]{Illingworth1977}
{Illingworth} G.,  1977, \mn@doi [\apjl] {10.1086/182572}, \href
  {http://adsabs.harvard.edu/abs/1977ApJ...218L..43I} {218, L43}

\bibitem[\protect\citeauthoryear{{Jesseit}, {Naab}  \& {Burkert}}{{Jesseit}
  et~al.}{2005}]{Jesseit2005}
{Jesseit} R.,  {Naab} T.,   {Burkert} A.,  2005, \mn@doi [\mnras]
  {10.1111/j.1365-2966.2005.09129.x}, \href
  {http://adsabs.harvard.edu/abs/2005MNRAS.360.1185J} {360, 1185}

\bibitem[\protect\citeauthoryear{{Jesseit}, {Cappellari}, {Naab}, {Emsellem}
  \& {Burkert}}{{Jesseit} et~al.}{2009}]{Jesseit2009}
{Jesseit} R.,  {Cappellari} M.,  {Naab} T.,  {Emsellem} E.,   {Burkert} A.,
  2009, \mn@doi [\mnras] {10.1111/j.1365-2966.2009.14984.x}, \href
  {http://adsabs.harvard.edu/abs/2009MNRAS.397.1202J} {397, 1202}

\bibitem[\protect\citeauthoryear{{Jing} \& {Suto}}{{Jing} \&
  {Suto}}{2002}]{Jing2002}
{Jing} Y.~P.,  {Suto} Y.,  2002, \mn@doi [\apj] {10.1086/341065}, \href
  {http://adsabs.harvard.edu/abs/2002ApJ...574..538J} {574, 538}

\bibitem[\protect\citeauthoryear{{Krajnovi{\'c}} et~al.,}{{Krajnovi{\'c}}
  et~al.}{2011}]{Krajnovic2011}
{Krajnovi{\'c}} D.,  et~al., 2011, \mn@doi [\mnras]
  {10.1111/j.1365-2966.2011.18560.x}, \href
  {http://adsabs.harvard.edu/abs/2011MNRAS.414.2923K} {414, 2923}

\bibitem[\protect\citeauthoryear{{Li}, {Li}, {Mao}, {Xu}, {Long}  \&
  {Emsellem}}{{Li} et~al.}{2016}]{Li2016}
{Li} H.,  {Li} R.,  {Mao} S.,  {Xu} D.,  {Long} R.~J.,   {Emsellem} E.,  2016,
  \mn@doi [\mnras] {10.1093/mnras/stv2565}, \href
  {http://adsabs.harvard.edu/abs/2016MNRAS.455.3680L} {455, 3680}

\bibitem[\protect\citeauthoryear{{Li} et~al.,}{{Li} et~al.}{2017}]{Li2017}
{Li} H.,  et~al., 2017, \mn@doi [\apj] {10.3847/1538-4357/aa662a}, \href
  {http://adsabs.harvard.edu/abs/2017ApJ...838...77L} {838, 77}

\bibitem[\protect\citeauthoryear{{Ma}, {Greene}, {McConnell}, {Janish},
  {Blakeslee}, {Thomas}  \& {Murphy}}{{Ma} et~al.}{2014}]{Ma2014}
{Ma} C.-P.,  {Greene} J.~E.,  {McConnell} N.,  {Janish} R.,  {Blakeslee} J.~P.,
   {Thomas} J.,   {Murphy} J.~D.,  2014, \mn@doi [\apj]
  {10.1088/0004-637X/795/2/158}, \href
  {http://adsabs.harvard.edu/abs/2014ApJ...795..158M} {795, 158}

\bibitem[\protect\citeauthoryear{{Martizzi}, {Teyssier}, {Moore}  \&
  {Wentz}}{{Martizzi} et~al.}{2012}]{Martizzi2012}
{Martizzi} D.,  {Teyssier} R.,  {Moore} B.,   {Wentz} T.,  2012, \mn@doi
  [\mnras] {10.1111/j.1365-2966.2012.20879.x}, \href
  {http://adsabs.harvard.edu/abs/2012MNRAS.422.3081M} {422, 3081}

\bibitem[\protect\citeauthoryear{{Monnet}, {Bacon}  \& {Emsellem}}{{Monnet}
  et~al.}{1992}]{Monnet1992}
{Monnet} G.,  {Bacon} R.,   {Emsellem} E.,  1992, \aap, \href
  {http://adsabs.harvard.edu/abs/1992A%26A...253..366M} {253, 366}

\bibitem[\protect\citeauthoryear{{Moody}, {Romanowsky}, {Cox}, {Novak}  \&
  {Primack}}{{Moody} et~al.}{2014}]{Moody2014}
{Moody} C.~E.,  {Romanowsky} A.~J.,  {Cox} T.~J.,  {Novak} G.~S.,   {Primack}
  J.~R.,  2014, \mn@doi [\mnras] {10.1093/mnras/stu1444}, \href
  {http://adsabs.harvard.edu/abs/2014MNRAS.444.1475M} {444, 1475}

\bibitem[\protect\citeauthoryear{{Naab} \& {Burkert}}{{Naab} \&
  {Burkert}}{2003}]{Naab2003}
{Naab} T.,  {Burkert} A.,  2003, \mn@doi [\apj] {10.1086/378581}, \href
  {http://adsabs.harvard.edu/abs/2003ApJ...597..893N} {597, 893}

\bibitem[\protect\citeauthoryear{{Naab} et~al.,}{{Naab}
  et~al.}{2014}]{Naab2014}
{Naab} T.,  et~al., 2014, \mn@doi [\mnras] {10.1093/mnras/stt1919}, \href
  {http://adsabs.harvard.edu/abs/2014MNRAS.444.3357N} {444, 3357}

\bibitem[\protect\citeauthoryear{{Navarro}, {Frenk}  \& {White}}{{Navarro}
  et~al.}{1996}]{Navarro1996}
{Navarro} J.~F.,  {Frenk} C.~S.,   {White} S.~D.~M.,  1996, \mn@doi [\apj]
  {10.1086/177173}, \href {http://adsabs.harvard.edu/abs/1996ApJ...462..563N}
  {462, 563}

\bibitem[\protect\citeauthoryear{{Nelson} et~al.,}{{Nelson}
  et~al.}{2015}]{Nelson2015}
{Nelson} D.,  et~al., 2015, \mn@doi [Astronomy and Computing]
  {10.1016/j.ascom.2015.09.003}, \href
  {http://adsabs.harvard.edu/abs/2015A%26C....13...12N} {13, 12}

\bibitem[\protect\citeauthoryear{{Novak}, {Cox}, {Primack}, {Jonsson}  \&
  {Dekel}}{{Novak} et~al.}{2006}]{Novak2006}
{Novak} G.~S.,  {Cox} T.~J.,  {Primack} J.~R.,  {Jonsson} P.,   {Dekel} A.,
  2006, \mn@doi [\apjl] {10.1086/506605}, \href
  {http://adsabs.harvard.edu/abs/2006ApJ...646L...9N} {646, L9}

\bibitem[\protect\citeauthoryear{{Penoyre}, {Moster}, {Sijacki}  \&
  {Genel}}{{Penoyre} et~al.}{2017}]{Penoyre2017}
{Penoyre} Z.,  {Moster} B.~P.,  {Sijacki} D.,   {Genel} S.,  2017, \mn@doi
  [\mnras] {10.1093/mnras/stx762}, \href
  {http://adsabs.harvard.edu/abs/2017MNRAS.468.3883P} {468, 3883}

\bibitem[\protect\citeauthoryear{{Pillepich} et~al.,}{{Pillepich}
  et~al.}{2017}]{Pillepich2017}
{Pillepich} A.,  et~al., 2017, preprint, \href
  {http://adsabs.harvard.edu/abs/2017arXiv170302970P} {} (\mn@eprint {arXiv}
  {1703.02970})

\bibitem[\protect\citeauthoryear{{Remus}, {Dolag}, {Naab}, {Burkert},
  {Hirschmann}, {Hoffmann}  \& {Johansson}}{{Remus} et~al.}{2017}]{Remus2017}
{Remus} R.-S.,  {Dolag} K.,  {Naab} T.,  {Burkert} A.,  {Hirschmann} M.,
  {Hoffmann} T.~L.,   {Johansson} P.~H.,  2017, \mn@doi [\mnras]
  {10.1093/mnras/stw2594}, \href
  {http://adsabs.harvard.edu/abs/2017MNRAS.464.3742R} {464, 3742}

\bibitem[\protect\citeauthoryear{{Rodriguez-Gomez} et~al.,}{{Rodriguez-Gomez}
  et~al.}{2015}]{Rodriguez-Gomez2015}
{Rodriguez-Gomez} V.,  et~al., 2015, \mn@doi [\mnras] {10.1093/mnras/stv264},
  \href {http://adsabs.harvard.edu/abs/2015MNRAS.449...49R} {449, 49}

\bibitem[\protect\citeauthoryear{{R{\"o}ttgers}, {Naab}  \&
  {Oser}}{{R{\"o}ttgers} et~al.}{2014}]{Rottgers2014}
{R{\"o}ttgers} B.,  {Naab} T.,   {Oser} L.,  2014, \mn@doi [\mnras]
  {10.1093/mnras/stu1762}, \href
  {http://adsabs.harvard.edu/abs/2014MNRAS.445.1065R} {445, 1065}

\bibitem[\protect\citeauthoryear{{Rybicki}}{{Rybicki}}{1987}]{Rybicki1987}
{Rybicki} G.~B.,  1987, in {de Zeeuw} P.~T.,  ed.,  IAU Symposium Vol. 127,
  Structure and Dynamics of Elliptical Galaxies. p.~397

\bibitem[\protect\citeauthoryear{{Ryden}}{{Ryden}}{1992}]{Ryden1992}
{Ryden} B.,  1992, \mn@doi [\apj] {10.1086/171731}, \href
  {http://adsabs.harvard.edu/abs/1992ApJ...396..445R} {396, 445}

\bibitem[\protect\citeauthoryear{{S{\'a}nchez} et~al.,}{{S{\'a}nchez}
  et~al.}{2012}]{Sanchez2012}
{S{\'a}nchez} S.~F.,  et~al., 2012, \mn@doi [\aap]
  {10.1051/0004-6361/201117353}, \href
  {http://adsabs.harvard.edu/abs/2012A%26A...538A...8S} {538, A8}

\bibitem[\protect\citeauthoryear{{Schaller} et~al.,}{{Schaller}
  et~al.}{2015}]{Schaller2015}
{Schaller} M.,  et~al., 2015, \mn@doi [\mnras] {10.1093/mnras/stv1341}, \href
  {http://adsabs.harvard.edu/abs/2015MNRAS.452..343S} {452, 343}

\bibitem[\protect\citeauthoryear{{Schaye} et~al.,}{{Schaye}
  et~al.}{2015}]{Schaye2015}
{Schaye} J.,  et~al., 2015, \mn@doi [\mnras] {10.1093/mnras/stu2058}, \href
  {http://adsabs.harvard.edu/abs/2015MNRAS.446..521S} {446, 521}

\bibitem[\protect\citeauthoryear{{S{\'e}rsic}}{{S{\'e}rsic}}{1963}]{sersic1963}
{S{\'e}rsic} J.~L.,  1963, Boletin de la Asociacion Argentina de Astronomia La
  Plata Argentina, \href {http://adsabs.harvard.edu/abs/1963BAAA....6...41S}
  {6, 41}

\bibitem[\protect\citeauthoryear{{Shi}, {Wang}  \& {Mo}}{{Shi}
  et~al.}{2015}]{Shi2015}
{Shi} J.,  {Wang} H.,   {Mo} H.~J.,  2015, \mn@doi [\apj]
  {10.1088/0004-637X/807/1/37}, \href
  {http://adsabs.harvard.edu/abs/2015ApJ...807...37S} {807, 37}

\bibitem[\protect\citeauthoryear{{Snyder} et~al.,}{{Snyder}
  et~al.}{2015}]{Snyder2015}
{Snyder} G.~F.,  et~al., 2015, \mn@doi [\mnras] {10.1093/mnras/stv2078}, \href
  {http://adsabs.harvard.edu/abs/2015MNRAS.454.1886S} {454, 1886}

\bibitem[\protect\citeauthoryear{{Springel}}{{Springel}}{2010}]{springel2010}
{Springel} V.,  2010, \mn@doi [\mnras] {10.1111/j.1365-2966.2009.15715.x},
  \href {http://adsabs.harvard.edu/abs/2010MNRAS.401..791S} {401, 791}

\bibitem[\protect\citeauthoryear{{Springel}, {White}, {Tormen}  \&
  {Kauffmann}}{{Springel} et~al.}{2001}]{Springel2001}
{Springel} V.,  {White} S.~D.~M.,  {Tormen} G.,   {Kauffmann} G.,  2001,
  \mn@doi [\mnras] {10.1046/j.1365-8711.2001.04912.x}, \href
  {http://adsabs.harvard.edu/abs/2001MNRAS.328..726S} {328, 726}

\bibitem[\protect\citeauthoryear{{Statler} \& {Fry}}{{Statler} \&
  {Fry}}{1994}]{Statler1994}
{Statler} T.~S.,  {Fry} A.~M.,  1994, \mn@doi [\apj] {10.1086/174002}, \href
  {http://adsabs.harvard.edu/abs/1994ApJ...425..481S} {425, 481}

\bibitem[\protect\citeauthoryear{{Tremblay} \& {Merritt}}{{Tremblay} \&
  {Merritt}}{1996}]{Tremblay1996}
{Tremblay} B.,  {Merritt} D.,  1996, \mn@doi [\aj] {10.1086/117959}, \href
  {http://adsabs.harvard.edu/abs/1996AJ....111.2243T} {111, 2243}

\bibitem[\protect\citeauthoryear{{Tsatsi}, {Lyubenova}, {van de Ven}, {Chang},
  {Aguerri}, {Falc{\'o}n-Barroso}  \& {Macci{\`o}}}{{Tsatsi}
  et~al.}{2017}]{Tsatsi2017}
{Tsatsi} A.,  {Lyubenova} M.,  {van de Ven} G.,  {Chang} J.,  {Aguerri}
  J.~A.~L.,  {Falc{\'o}n-Barroso} J.,   {Macci{\`o}} A.~V.,  2017, preprint,
  \href {http://adsabs.harvard.edu/abs/2017arXiv170705130T} {} (\mn@eprint
  {arXiv} {1707.05130})

\bibitem[\protect\citeauthoryear{{Veale}, {Ma}, {Greene}, {Thomas},
  {Blakeslee}, {McConnell}, {Walsh}  \& {Ito}}{{Veale}
  et~al.}{2017a}]{Veale2017b}
{Veale} M.,  {Ma} C.-P.,  {Greene} J.~E.,  {Thomas} J.,  {Blakeslee} J.~P.,
  {McConnell} N.,  {Walsh} J.~L.,   {Ito} J.,  2017a, preprint, \href
  {http://adsabs.harvard.edu/abs/2017arXiv170308573V} {} (\mn@eprint {arXiv}
  {1703.08573})

\bibitem[\protect\citeauthoryear{{Veale} et~al.,}{{Veale}
  et~al.}{2017b}]{Veale2017a}
{Veale} M.,  et~al., 2017b, \mn@doi [\mnras] {10.1093/mnras/stw2330}, \href
  {http://adsabs.harvard.edu/abs/2017MNRAS.464..356V} {464, 356}

\bibitem[\protect\citeauthoryear{{Velliscig} et~al.,}{{Velliscig}
  et~al.}{2015}]{Velliscig2015}
{Velliscig} M.,  et~al., 2015, \mn@doi [\mnras] {10.1093/mnras/stv1690}, \href
  {http://adsabs.harvard.edu/abs/2015MNRAS.453..721V} {453, 721}

\bibitem[\protect\citeauthoryear{{Vogelsberger}, {Genel}, {Sijacki}, {Torrey},
  {Springel}  \& {Hernquist}}{{Vogelsberger} et~al.}{2013}]{Vogelsberger2013}
{Vogelsberger} M.,  {Genel} S.,  {Sijacki} D.,  {Torrey} P.,  {Springel} V.,
  {Hernquist} L.,  2013, \mn@doi [\mnras] {10.1093/mnras/stt1789}, \href
  {http://adsabs.harvard.edu/abs/2013MNRAS.436.3031V} {436, 3031}

\bibitem[\protect\citeauthoryear{{Vogelsberger} et~al.,}{{Vogelsberger}
  et~al.}{2014a}]{Vogelsberger2014a}
{Vogelsberger} M.,  et~al., 2014a, \mn@doi [\mnras] {10.1093/mnras/stu1536},
  \href {http://adsabs.harvard.edu/abs/2014MNRAS.444.1518V} {444, 1518}

\bibitem[\protect\citeauthoryear{{Vogelsberger} et~al.,}{{Vogelsberger}
  et~al.}{2014b}]{Vogelsberger2014b}
{Vogelsberger} M.,  et~al., 2014b, \mn@doi [\nat] {10.1038/nature13316}, \href
  {http://adsabs.harvard.edu/abs/2014Natur.509..177V} {509, 177}

\bibitem[\protect\citeauthoryear{{Weijmans} et~al.,}{{Weijmans}
  et~al.}{2014}]{Weijmans2014}
{Weijmans} A.-M.,  et~al., 2014, \mn@doi [\mnras] {10.1093/mnras/stu1603},
  \href {http://adsabs.harvard.edu/abs/2014MNRAS.444.3340W} {444, 3340}

\bibitem[\protect\citeauthoryear{{Weinberger} et~al.,}{{Weinberger}
  et~al.}{2017}]{Weinberger2017}
{Weinberger} R.,  et~al., 2017, \mn@doi [\mnras] {10.1093/mnras/stw2944}, \href
  {http://adsabs.harvard.edu/abs/2017MNRAS.465.3291W} {465, 3291}

\bibitem[\protect\citeauthoryear{{Welker}, {Devriendt}, {Dubois}, {Pichon}  \&
  {Peirani}}{{Welker} et~al.}{2014}]{Welker2014}
{Welker} C.,  {Devriendt} J.,  {Dubois} Y.,  {Pichon} C.,   {Peirani} S.,
  2014, \mn@doi [\mnras] {10.1093/mnrasl/slu106}, \href
  {http://adsabs.harvard.edu/abs/2014MNRAS.445L..46W} {445, L46}

\bibitem[\protect\citeauthoryear{{Xu}, {Springel}, {Sluse}, {Schneider},
  {Sonnenfeld}, {Nelson}, {Vogelsberger}  \& {Hernquist}}{{Xu}
  et~al.}{2017}]{Xu2017}
{Xu} D.,  {Springel} V.,  {Sluse} D.,  {Schneider} P.,  {Sonnenfeld} A.,
  {Nelson} D.,  {Vogelsberger} M.,   {Hernquist} L.,  2017, \mn@doi [\mnras]
  {10.1093/mnras/stx899}, \href
  {http://adsabs.harvard.edu/abs/2017MNRAS.469.1824X} {469, 1824}

\bibitem[\protect\citeauthoryear{{Zhang}, {Yang}, {Wang}, {Wang}, {Mo}  \& {van
  den Bosch}}{{Zhang} et~al.}{2013}]{Zhang2013}
{Zhang} Y.,  {Yang} X.,  {Wang} H.,  {Wang} L.,  {Mo} H.~J.,   {van den Bosch}
  F.~C.,  2013, \mn@doi [\apj] {10.1088/0004-637X/779/2/160}, \href
  {http://adsabs.harvard.edu/abs/2013ApJ...779..160Z} {779, 160}

\bibitem[\protect\citeauthoryear{{Zheng} et~al.,}{{Zheng}
  et~al.}{2017}]{Zheng2017}
{Zheng} Z.,  et~al., 2017, \mn@doi [\mnras] {10.1093/mnras/stw3030}, \href
  {http://adsabs.harvard.edu/abs/2017MNRAS.465.4572Z} {465, 4572}

\bibitem[\protect\citeauthoryear{{van den Bosch}}{{van den
  Bosch}}{1997}]{Bosch1997}
{van den Bosch} F.~C.,  1997, \mn@doi [\mnras] {10.1093/mnras/287.3.543}, \href
  {http://adsabs.harvard.edu/abs/1997MNRAS.287..543V} {287, 543}

\makeatother
\end{thebibliography}



\appendix

\section{Images, velocity maps and merger trees for all the prolate galaxies in the sample}
\label{AppendixA}
We present here the images, velocity maps and the merger history of all the prolate galaxies
in our sample. The merger trees for subhalo0 and subhalo66080 break in the middle. This is 
due to a small technical problem in the {\small SUBLINK} tree, which, however, has no effects on our conclusion.
\begin{figure*}
\includegraphics[width=\textwidth]{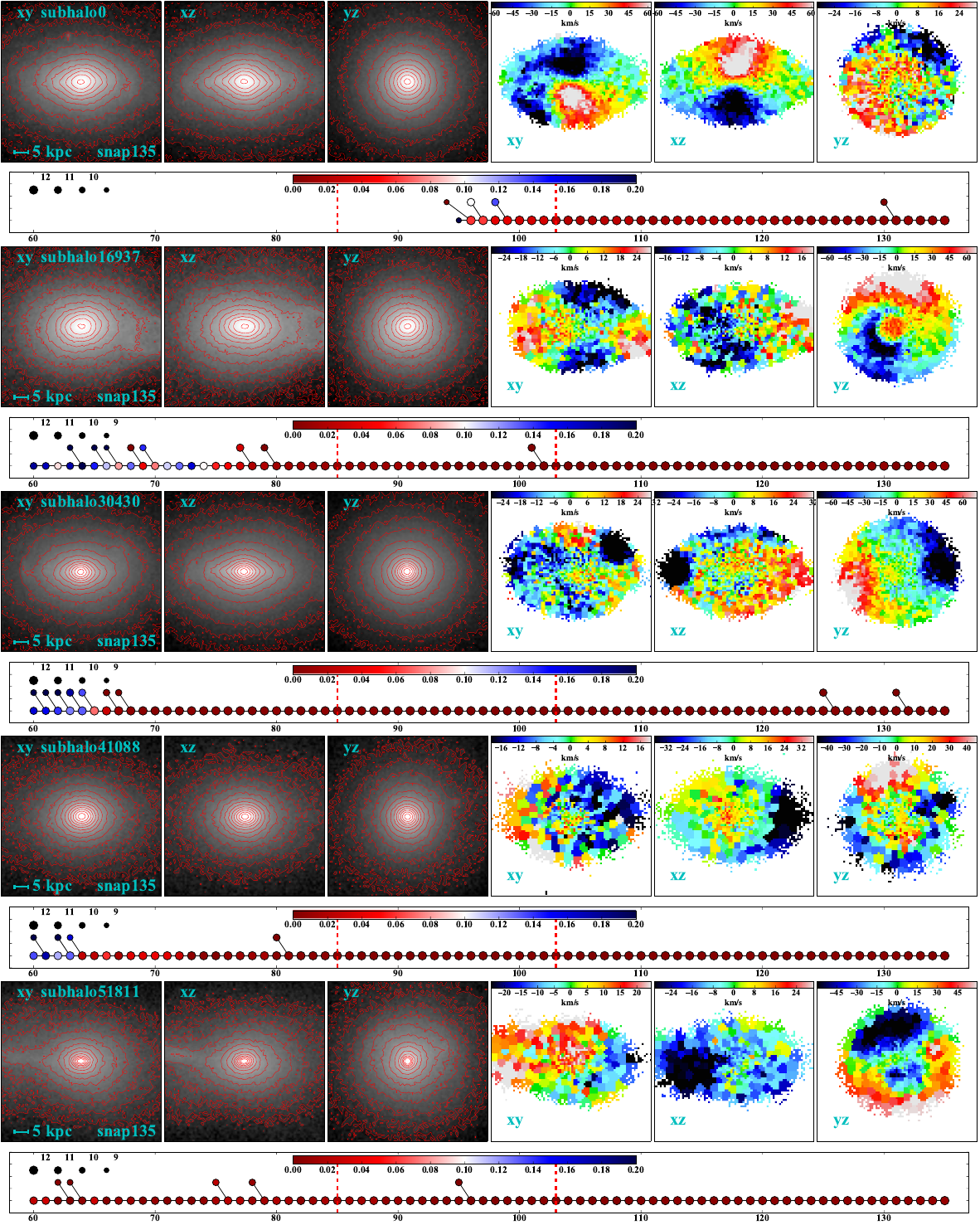}
   \caption{Images, isophotes, line-of-sight velocity maps and merger trees for all the prolate
    galaxies in our sample. The labels corresponds to those in Fig.~\ref{fig:subhalo210738}.}
   \label{fig:A1}    
\end{figure*}

\begin{figure*}
\includegraphics[width=\textwidth]{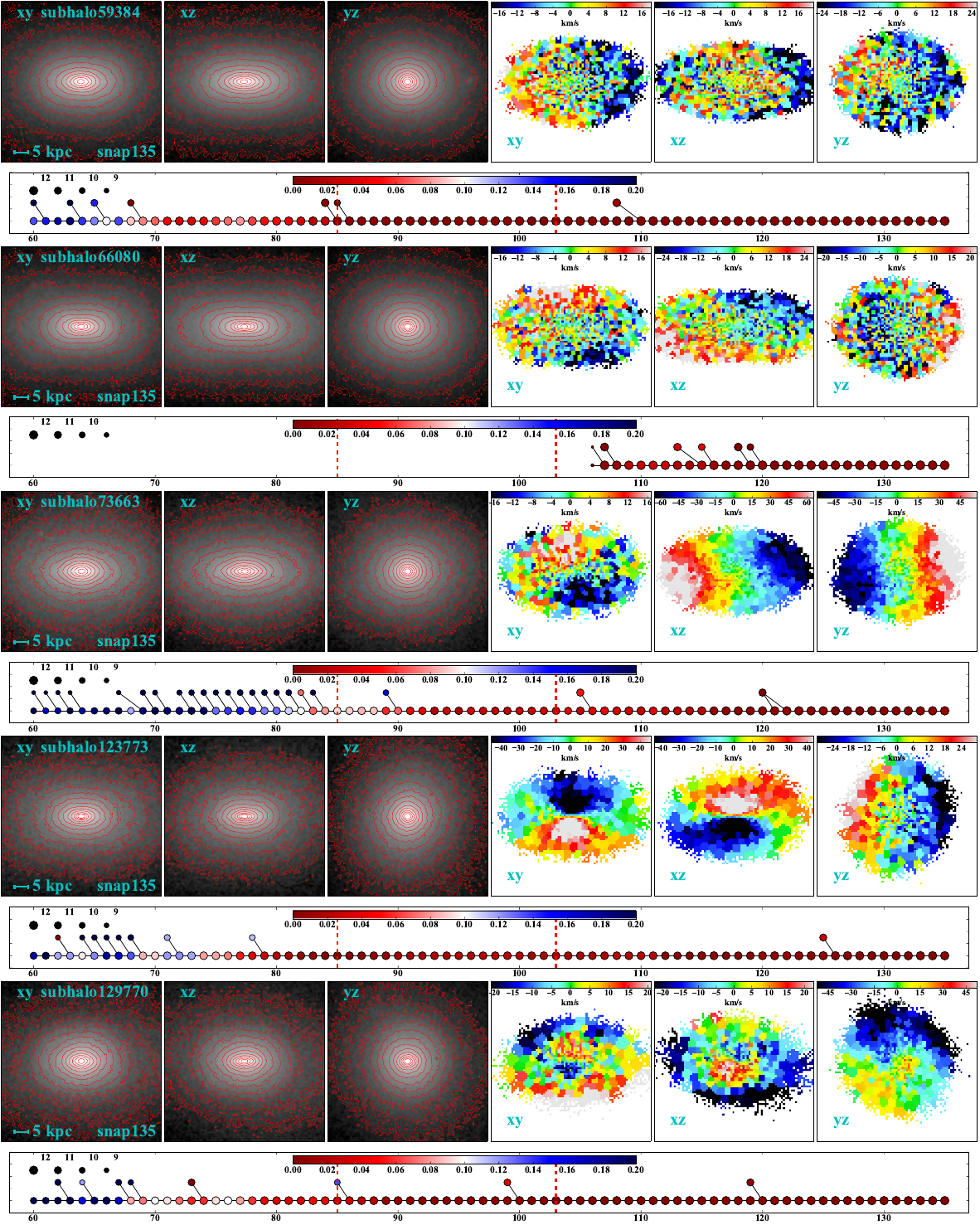}
\addtocounter{figure}{-1}
   \caption{--- continued}
\end{figure*}

\begin{figure*}
\includegraphics[width=\textwidth]{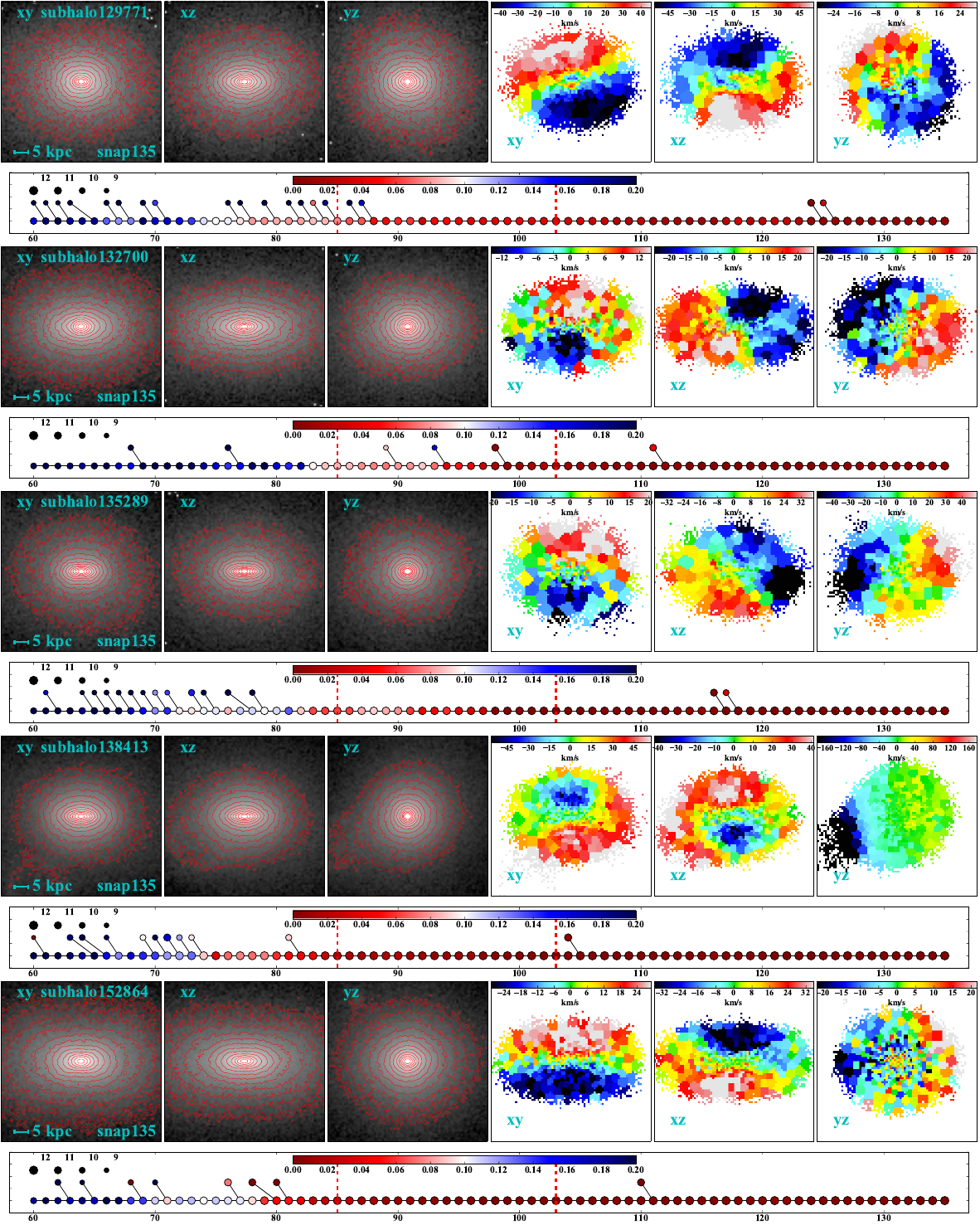}
\addtocounter{figure}{-1}
   \caption{--- continued}
\end{figure*}

\begin{figure*}
\includegraphics[width=\textwidth]{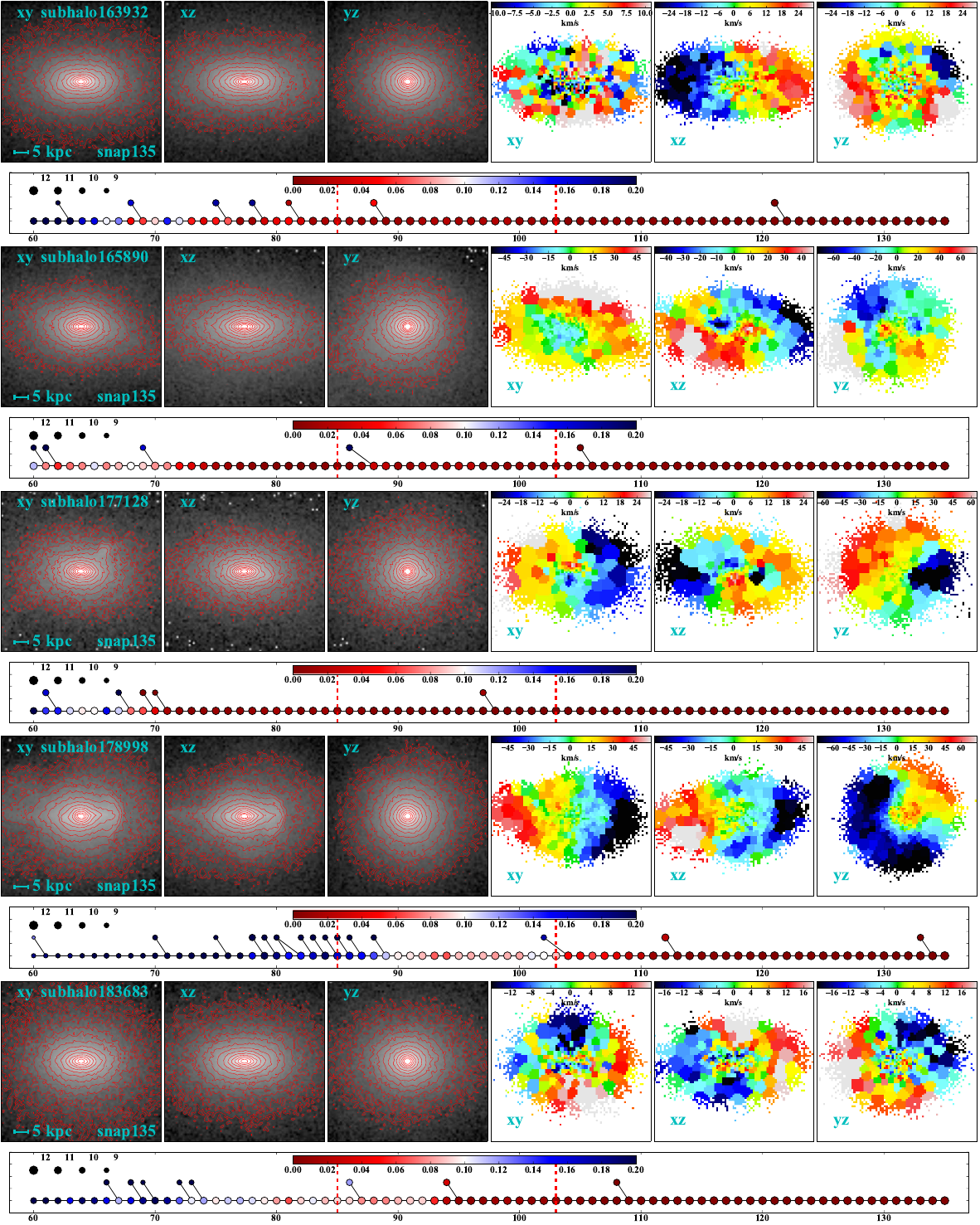}
\addtocounter{figure}{-1}
   \caption{--- continued}
\end{figure*}

\begin{figure*}
\includegraphics[width=\textwidth]{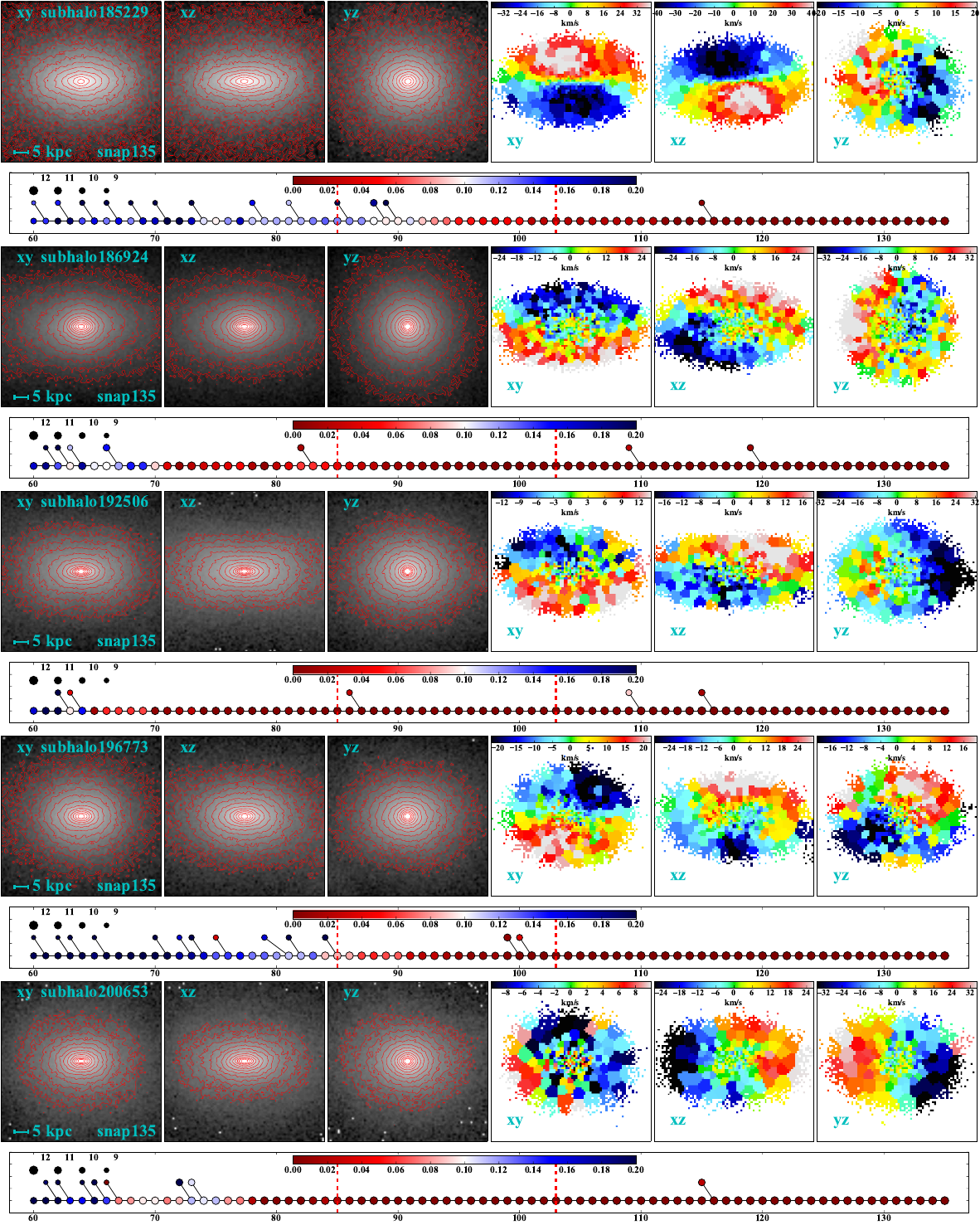}
\addtocounter{figure}{-1}
   \caption{--- continued}
\end{figure*}

\begin{figure*}
\includegraphics[width=\textwidth]{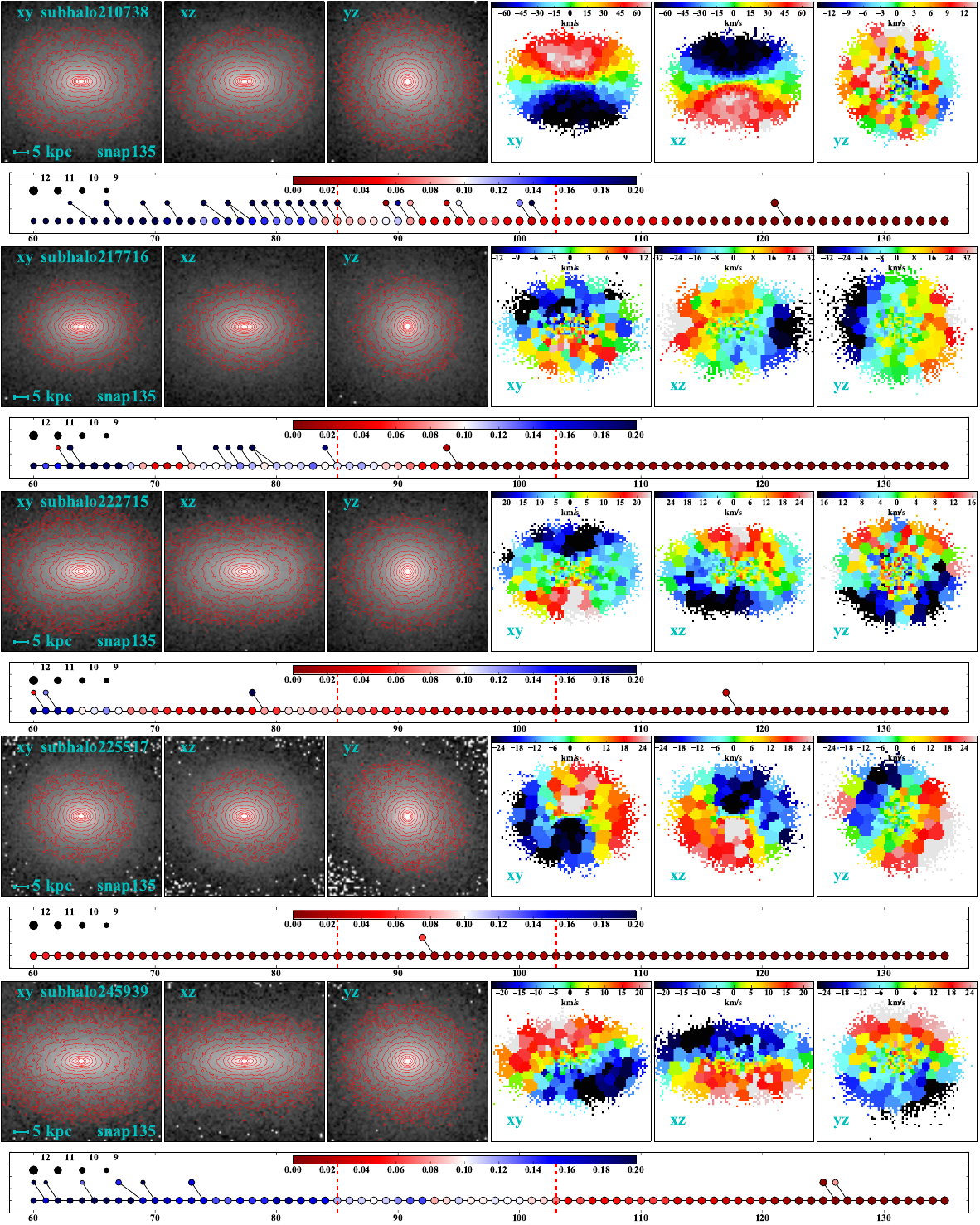}
\addtocounter{figure}{-1}
   \caption{--- continued}
\end{figure*}

\begin{figure*}
\includegraphics[width=\textwidth]{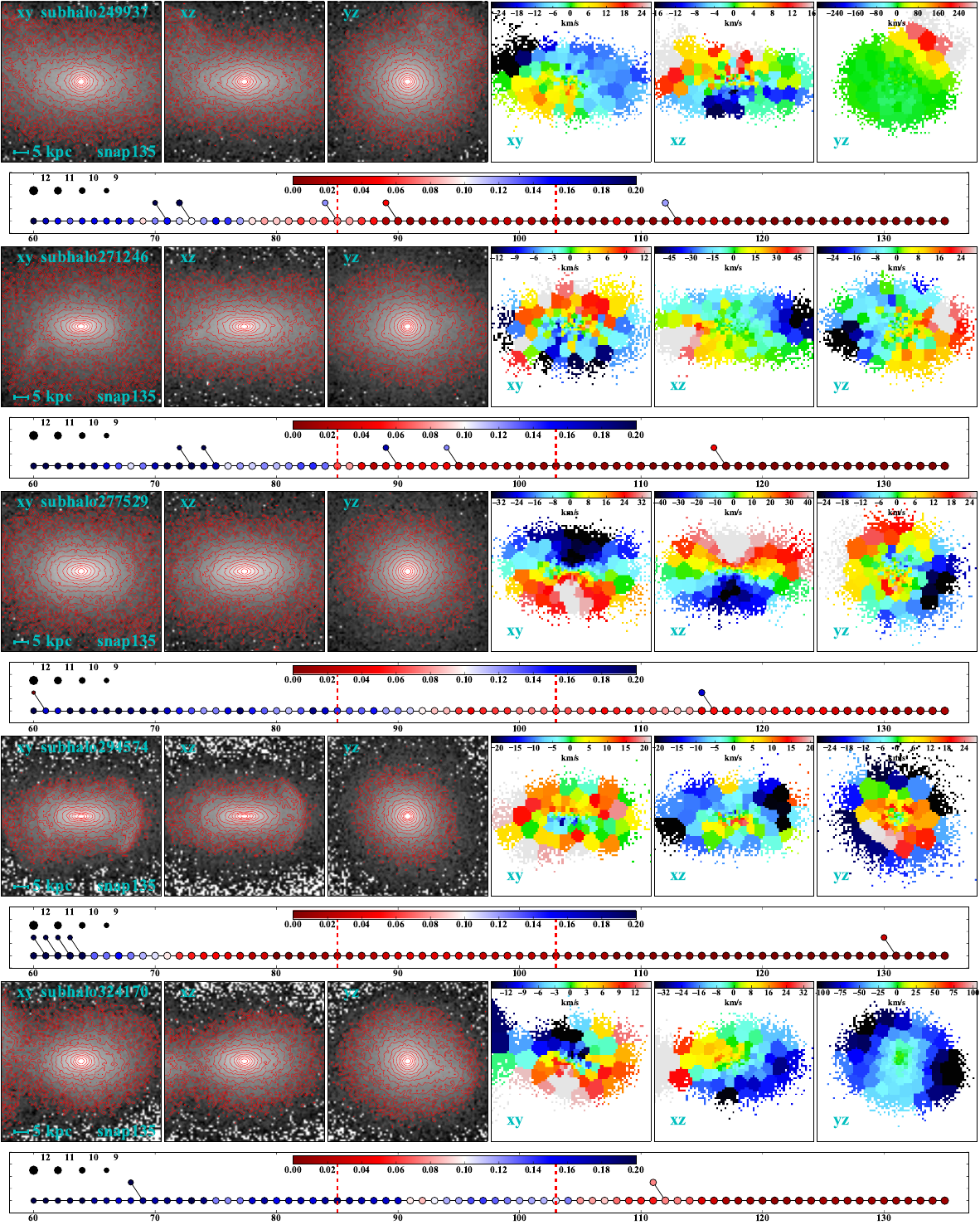}
\addtocounter{figure}{-1}
   \caption{--- continued}
\end{figure*}

\section{Properties of the prolate galaxies in the sample}
\clearpage
\begin{deluxetable}{lccccccccccccccc}
\tablewidth{0pt}
\tabletypesize{\small}
\setlength{\tabcolsep}{4pt}
\tablecaption{Properties of all the prolate galaxies in the sample\label{tab:properties}}
\tablehead{
 \colhead{subhaloID} &
 \colhead{$\log M^*$} &
 \colhead{$M_{\rm c200}$} &
 \colhead{$R_{\rm c200}$} &
 \colhead{$r^{*}_{\rm h}$} &
 \colhead{$b/a$} &
 \colhead{$c/a$} &
 \colhead{$\sigma_{\rm 0.5R_e}$} &
 \colhead{$\gamma^{*}$} &
 \colhead{$\gamma^{\rm D}$} &
 \colhead{$\gamma^{\rm T}$} & 
 \colhead{$f_{\rm DM}(<\rm R_e)$} & 
 \colhead{$\lambda_{\rm R_e}$} &
 \colhead{$\log R_e$} &
 \colhead{$\varepsilon$} &
 \colhead{$\beta$} \\
 \colhead{} &
 \colhead{($M_{\odot}$)} &
 \colhead{($M_{\odot}$)} &
 \colhead{(Mpc)} &
 \colhead{(kpc)} &
 \colhead{} &
 \colhead{} &
 \colhead{(km/s)} &
 \colhead{} &
 \colhead{} &
 \colhead{} &
 \colhead{} &
 \colhead{} &
 \colhead{(kpc)} &
 \colhead{} &
 \colhead{} \\
 \colhead{(1)} &
 \colhead{(2)} &
 \colhead{(3)} &
 \colhead{(4)} &
 \colhead{(5)} &
 \colhead{(6)} &
 \colhead{(7)} &
 \colhead{(8)} &
 \colhead{(9)} &
 \colhead{(10)} &
 \colhead{(11)} &
 \colhead{(12)} &
 \colhead{(13)} &
 \colhead{(14)} &
 \colhead{(15)} &
 \colhead{(16)}
}
\startdata
\bf{subhalo0     } & 12.52& 14.37& 1.263& 102.3&  0.63&  0.60& 343.8& -2.75& -1.32& -1.78&  0.64& 0.069&  1.62&  0.23&  0.28 \\
    subhalo16937   & 12.35& 14.35& 1.246& 138.3&  0.63&  0.56& 249.2& -2.86& -1.46& -1.72&  0.64& 0.049&  1.53&  0.43&  0.16 \\
    subhalo30430   & 12.49& 14.34& 1.237&  72.3&  0.68&  0.62& 346.5& -2.51& -1.19& -1.71&  0.58& 0.041&  1.53&  0.28&  0.34 \\
    subhalo41088   & 12.16& 14.07& 1.005&  88.5&  0.73&  0.71& 270.3& -2.96& -1.39& -1.64&  0.73& 0.047&  1.51&  0.34&  0.22 \\
    subhalo51811   & 12.08& 14.23& 1.137&  89.7&  0.55&  0.52& 233.1& -2.43& -1.24& -1.47&  0.75& 0.061&  1.53&  0.56&  0.31 \\
    subhalo59384   & 12.42& 14.11& 1.035&  54.6&  0.59&  0.49& 338.9& -2.56& -1.38& -1.90&  0.59& 0.038&  1.48&  0.30&  0.31 \\
\bf{subhalo66080 } & 12.42& 14.13& 1.057&  60.0&  0.53&  0.42& 349.1& -2.46& -1.28& -1.78&  0.59& 0.042&  1.50&  0.45&  0.32 \\
    subhalo73663   & 12.11& 13.64& 0.725&  37.9&  0.57&  0.47& 302.6& -2.66& -1.56& -2.15&  0.47& 0.051&  1.29&  0.21&  0.28 \\
\bf{subhalo123773} & 11.98& 13.77& 0.801&  39.9&  0.54&  0.52& 255.7& -2.58& -1.50& -1.96&  0.58& 0.086&  1.35&  0.27&  0.32 \\
    subhalo129770  & 12.02& 13.57& 0.684&  42.1&  0.69&  0.60& 229.9& -2.65& -1.50& -2.04&  0.55& 0.049&  1.37&  0.29&  0.32 \\
\bf{subhalo129771} & 11.68& 13.57& 0.684&  16.5&  0.69&  0.57& 212.7& -2.39& -1.26& -2.01&  0.44& 0.050&  1.09&  0.10&  0.25 \\
\bf{subhalo132700} & 11.76& 13.33& 0.570&  19.8&  0.61&  0.46& 201.2& -2.53& -1.27& -2.05&  0.50& 0.047&  1.18&  0.38&  0.35 \\
\bf{subhalo135289} & 11.78& 13.56& 0.682&  32.5&  0.52&  0.41& 219.1& -2.93& -1.66& -2.22&  0.52& 0.032&  1.21&  0.13&  0.40 \\
\bf{subhalo138413} & 12.03& 13.77& 0.801&  24.6&  0.51&  0.47& 279.6& -2.78& -1.22& -2.21&  0.41& 0.036&  1.15&  0.29&  0.32 \\
\bf{subhalo152864} & 11.94& 13.63& 0.716&  27.4&  0.58&  0.47& 286.9& -2.51& -1.33& -1.93&  0.57& 0.039&  1.22&  0.31&  0.30 \\
    subhalo163932  & 11.86& 13.46& 0.632&  28.3&  0.55&  0.46& 252.4& -2.58& -1.41& -1.97&  0.59& 0.037&  1.26&  0.31&  0.34 \\
    subhalo165890  & 11.80& 13.62& 0.713&  28.0&  0.53&  0.45& 271.9& -2.82& -1.43& -2.02&  0.62& 0.039&  1.20&  0.22&  0.23 \\
    subhalo177128  & 11.56& 13.41& 0.608&  33.3&  0.57&  0.52& 167.9& -2.90& -1.39& -1.91&  0.66& 0.053&  1.22&  0.33&  0.37 \\
    subhalo178998  & 11.85& 13.62& 0.712&  33.3&  0.59&  0.53& 218.5& -2.55& -1.48& -2.00&  0.54& 0.078&  1.30&  0.33&  0.35 \\
\bf{subhalo183683} & 11.68& 13.51& 0.653&  25.4&  0.58&  0.49& 221.2& -2.57& -1.53& -2.03&  0.57& 0.036&  1.19&  0.35&  0.39 \\
\bf{subhalo185229} & 11.67& 13.08& 0.471&  23.5&  0.52&  0.41& 199.4& -1.96& -1.14& -1.58&  0.57& 0.117&  1.22&  0.52&  0.30 \\
\bf{subhalo186924} & 12.06& 13.51& 0.655&  33.5&  0.66&  0.63& 283.7& -2.90& -1.50& -2.17&  0.59& 0.044&  1.35&  0.36&  0.25 \\
\bf{subhalo192506} & 11.78& 13.43& 0.614&  22.4&  0.49&  0.41& 217.9& -2.66& -1.30& -2.04&  0.55& 0.039&  1.19&  0.50&  0.38 \\
\bf{subhalo196773} & 11.78& 13.38& 0.594&  20.8&  0.62&  0.54& 237.1& -2.56& -1.38& -2.11&  0.47& 0.044&  1.14&  0.33&  0.35 \\
    subhalo200653  & 11.66& 13.34& 0.573&  16.0&  0.54&  0.44& 207.6& -2.58& -1.09& -2.07&  0.45& 0.029&  1.04&  0.42&  0.36 \\
\bf{subhalo210738} & 11.73& 13.14& 0.492&  18.4&  0.44&  0.44& 240.6& -2.55& -1.49& -2.15&  0.45& 0.111&  1.09&  0.20&  0.33 \\
    subhalo217716  & 11.68& 13.22& 0.523&  15.0&  0.46&  0.42& 223.9& -2.60& -1.23& -2.19&  0.39& 0.029&  1.00&  0.37&  0.35 \\
\bf{subhalo222715} & 11.72& 13.07& 0.466&  18.2&  0.55&  0.53& 257.4& -2.30& -1.17& -1.88&  0.44& 0.033&  1.03&  0.10&  0.34 \\
\bf{subhalo225517} & 11.50& 13.20& 0.515&  11.7&  0.56&  0.56& 203.5& -2.64& -1.18& -2.09&  0.51& 0.101&  0.94&  0.20&  0.31 \\
\bf{subhalo245939} & 11.57& 13.06& 0.465&  19.8&  0.49&  0.38& 234.4& -2.28& -1.27& -1.77&  0.54& 0.039&  1.05&  0.08&  0.36 \\
    subhalo249937  & 11.44& 13.05& 0.459&  24.0&  0.69&  0.61& 189.8& -2.59& -1.38& -1.95&  0.58& 0.059&  1.14&  0.25&  0.32 \\
    subhalo271246  & 11.47& 13.03& 0.454&  24.4&  0.55&  0.47& 170.6& -2.57& -1.49& -1.98&  0.62& 0.048&  1.20&  0.49&  0.31 \\
\bf{subhalo277529} & 11.44& 12.78& 0.374&  16.6&  0.56&  0.51& 191.1& -2.12& -1.29& -1.74&  0.54& 0.090&  1.05&  0.42&  0.28 \\
    subhalo294574  & 11.25& 12.82& 0.385&  13.9&  0.35&  0.31& 125.4& -2.32& -0.70& -1.78&  0.54& 0.055&  1.02&  0.55&  0.42 \\
    subhalo324170  & 11.28& 12.66& 0.342&  19.1&  0.64&  0.61& 155.3& -2.55& -1.21& -1.91&  0.54& 0.060&  1.03&  0.06&  0.09 \\
\enddata
\tablecomments{
Column (1): The SUBFIND ID at snapshot 135. The galaxies shown with bold font have minor-axis rotation.
Column (2): Total stellar mass.
Column (3): Total mass enclosed in a sphere whose mean density is 200 times the critical density of the Universe.
Column (4): Radius of a sphere whose mean density is 200 times the critical density of the Universe. 
Column (5): Half stellar mass radius.
Column (6): Axis ratio $b/a$ calculated within the half stellar mass radius.
Column (7): Axis ratio $c/a$ calculated within the half stellar mass radius. 
Column (8): Velocity dispersion within 0.5 $R_e$. 
Column (9): Average stellar density slope between $0.1r^{*}_{\rm h}$ and $0.5r^{*}_{\rm h}$.
Column (10): Average dark matter density slope between $0.1r^{*}_{\rm h}$ and $0.5r^{*}_{\rm h}$. 
Column (11): Average total mass density slope between $0.1r^{*}_{\rm h}$ and $0.5r^{*}_{\rm h}$.
Column (12): Dark matter fraction within $R_e$. 
Column (13): Parameter $\lambda_{\rm R_e}$. 
Column (14): Effective radius.
Column (15): Average ellipticity within $R_e$. 
Column (16): Velocity anisotropy parameter $\beta$ within $R_e$.
The information in columns (2), (3), (4) and (5) is described in \citet{Nelson2015}, the information
in column (16) is described in \citet{Xu2017}.
}
\end{deluxetable}

\bsp	
\label{lastpage}
\end{document}